\documentclass[pra,12pt,tightenlines,superscriptaddress,nofootinbib,eqsecnum,secnumarabic]{revtex4}

\usepackage{bm,amssymb,amsmath,amsthm,mathrsfs,enumerate,verbatim}

\theoremstyle{plain}
\newtheorem{thm}{Theorem}[section]
\newtheorem{lem}[thm]{Lemma}
\newtheorem{cor}[thm]{Corollary}
\newtheorem{prp}[thm]{Proposition}

\theoremstyle{definition}
\newtheorem{dfn}[thm]{Definition}

\theoremstyle{remark}

\DeclareSymbolFont{AMSb}{U}{msb}{m}{n}
\DeclareMathSymbol{\N}{\mathbin}{AMSb}{"4E}
\DeclareMathSymbol{\Z}{\mathbin}{AMSb}{"5A}
\DeclareMathSymbol{\R}{\mathbin}{AMSb}{"52}
\DeclareMathSymbol{\Q}{\mathbin}{AMSb}{"51}
\DeclareMathSymbol{\I}{\mathbin}{AMSb}{"49}
\DeclareMathSymbol{\C}{\mathbin}{AMSb}{"43}
\DeclareMathSymbol{\F}{\mathbin}{AMSb}{"46}
\DeclareMathSymbol{\E}{\mathbin}{AMSb}{"45}

\DeclareSymbolFont{symbolsC}{U}{txsyc}{m}{n}
\DeclareMathSymbol{\coloneq}{\mathrel}{symbolsC}{66}

\def\tr{\operatorname{tr}}
\def\ket#1{|#1\rangle}
\def\bra#1{\langle#1|}
\def\ketbra#1{| #1 \rangle\langle #1 |}
\def\braket#1#2{\langle#1|#2\rangle}
\def\Ket#1{|#1)}
\def\Bra#1{(#1|}
\def\KetBra#1{|#1)(#1|}
\def\BraKet#1#2{(#1|#2)}
\def\Ketb#1{\big|#1\big)}
\def\Brab#1{\big(#1\big|}
\def\KetBrab#1{\big|#1\big)\big(#1\big|}
\def\BraKetb#1#2{\big(#1\big|#2\big)}

\def\End{\operatorname{End}}
\def\Hom{\operatorname{Hom}}
\def\Pisym{\Pi_\mathrm{sym}}

\def\d{\mathrm{d}}
\def\Qd{\operatorname{Q}(\C^d)}
\def\Q2{\operatorname{Q}(\C^2)}

\def\Is{\mathrm{\bf I}}
\def\Tr{\operatorname{Tr}}
\def\Hd{\operatorname{H}_0(\C^d)}
\def\muu{\mu}
\def\Pisymt{\Pi_\mathrm{sym}^{(t)}}
\def\Ptr{\bm{\Pi}_0}
\def\Ih{\mathrm{\bf I}_{\scriptscriptstyle{\mathrm{H}_0}}}

\def\ip{\braket}
\newcommand\abs[1]{\left|#1\right|}

\newcommand\om{\omega}
\newcommand\Om{\Omega}

\def\re{\R}

\begin{document}

\title{Weighted complex projective 2-designs from bases: \\ optimal state determination by orthogonal measurements}

\author{Aidan Roy}
\email{aroy@qis.ucalgary.ca}
\affiliation{Institute for Quantum Information Science, University of Calgary, Calgary, Alberta T2N 1N4, Canada}
\author{A. J. Scott}
\email{andrew.scott@griffith.edu.au}
\affiliation{Centre for Quantum Computer Technology, Centre for Quantum Dynamics, School of Science, Griffith University, Brisbane, Queensland 4111, Australia}

\begin{abstract}
We introduce the problem of constructing weighted complex projective 2-designs from the union of a family of orthonormal bases.
If the weight remains constant across elements of the same basis, then such designs can be interpreted as generalizations of 
complete sets of mutually unbiased bases, being equivalent whenever the design is composed of $d+1$ bases in dimension $d$. 
We show that, for the purpose of quantum state determination, these designs specify an optimal collection of orthogonal measurements. 
Using highly nonlinear functions on abelian groups, we construct explicit examples from $d+2$ orthonormal bases whenever $d+1$ is a prime power, covering dimensions $d=6$, 10, and 12, for example, 
where no complete sets of mutually unbiased bases have thus far been found.
\end{abstract}


\maketitle

\section{Introduction}
\label{sec:intro}

Fundamental to the fabrication of quantum information processing devices~\cite{Nielsen00}, such as quantum teleporters, 
key distributers, cloners, gates, and indeed, quantum computers, is the ability to precisely determine an unknown quantum state. 
Quality assurance requires a complete characterization of these devices, which can be accomplished through knowledge of 
the output states for a judicious choice of input states.

The determination of an unknown state of a quantum system is achieved by a sequence of measurements on 
identically prepared copies of the system. If the outcome statistics for the measurements uniquely identify each member 
from the set of all quantum states, then an estimate of these statistics will reveal the particular state under 
examination. This process is called {\em quantum state tomography}~\cite{Paris04}. 

One method of tomography is to perform identical measurements on each system copy; in this case, complete state determination requires 
that the outcome statistics for this single repeated measurement be described by an informationally complete positive-operator-valued 
measure (IC-POVM)~\cite{Renes04,Scott06}. In the original tomographic paradigm~\cite{Fano57,Park71,Ivanovic83,Leonhardt95}, 
however, the measurements differ: each one is orthogonal and prescribed by a member of an informationally complete set of quantum 
observables~\cite{Prugovecki77,Schroeck89,Busch89,Busch91}. The standard example for this latter scenario is provided by a complete 
set of mutually unbiased bases (MUBs)~\cite{Ivanovic81,Wootters89}. That is, a maximal set of $d+1$ orthonormal bases, 
$\{\ket{e^a_j}\}_{j=0}^{d-1}\subset\C^d$, $a=0,\dots,d$, having a constant overlap of $1/d$ between elements of different bases:
\begin{equation}
|\braket{e_j^a}{e_k^b}|^2 \;=\; \begin{cases}
\delta_{jk}\,, &\; a=b\,; \\ 
1/d\,, &\; a\neq b\;.
\end{cases}
\end{equation}
In dimension 2, for example, the 3 bases correspond to ``spin'' measurements along the $x$, $y$, and $z$ axes of the Bloch sphere. 

Explicit constructions of complete sets of MUBs are known for all prime-power dimensions $d=p^n$~\cite{Alltop80,Ivanovic81,Wootters89,Calderbank97,Bandyopadhyay02,Klappenecker04,Wocjan05,Planat06,Godsil05}. 
There is currently no supporting evidence, however, for their existence in other dimensions~\cite{Saniga04,Bengtsson05,Archer05,Wootters06,Aschbacher07,Boykin05}. 
Indeed, even in dimension 6, only sets of 3 MUBs have thus far been found~\cite{Grassl04,Bengtsson06,Butterley07}, which falls well 
short of the 7 needed for the complete determination of a quantum state in this dimension. It is thus timely to search for 
alternative sets of bases that are also suitable for quantum state determination, but retain important properties of MUBs in this 
role.

In this article we investigate weighted complex projective 2-designs that are formed by the union of a family of orthonormal bases. General 
complex projective $t$-designs and their variants have recently attracted attention from the perspective of quantum information theory~\cite{Zauner99,Barnum00,Renes04,Klappenecker05b,Hayashi05,Scott06,Ballester05,Dankert06,Gross06,Kim06,Ambainis07}.
Within the context of quantum state determination, weighted \mbox{2-designs} in $\C P^{d-1}$ that are formed by the union of a family 
of orthonormal bases for $\C^d$ can be interpreted as generalizations of complete sets of MUBs, provided that the weight remains 
constant across elements of the same basis. In such cases, they specify a series of orthogonal measurements whose outcome statistics 
are together described by a tight IC-POVM~\cite{Scott06}. The smallest number of bases that can be used to construct a weighted 
$2$-design in $\C P^{d-1}$ is $d+1$; when exactly $d+1$ bases are used, these designs are equivalent to complete sets of MUBs. If an additional basis is allowed, however, we find that weighted 2-designs can be constructed from a 
family of $d+2$ orthonormal bases whenever $d+1$ is a prime power, covering dimensions $d=6$, 10, and 12, for example, where no complete sets of MUBs have thus far been found. Explicitly, in dimension 6, by appending the standard basis 
$\{\ket{e^0_j}\coloneq\ket{e_j}\}_{j=0}^{5}$ to the 7 bases with elements
\begin{equation}
\ket{e_j^a} \;\coloneq\; \frac{1}{\sqrt{6}} \sum_{k=0}^{5} e^{2\pi i jk/6}e^{2\pi i a 3^k/7}\ket{e_k} \qquad (a=1,\dots,7)\;,
\end{equation}
we obtain a family of 8 orthonormal bases whose union forms a weighted complex projective 2-design. All members of the standard basis 
are appointed the weight $w_0=1/42$, while all members of the remaining bases are appointed the weight $w_a=1/49$. 

Most importantly, returning to the task of quantum state tomography, by measuring copies of a quantum system in the standard basis at 
a frequency ratio of $7:6$ relative to each other basis, we retain the same minimal error rate in our estimate of the system state as 
that for a complete set of MUBs (if one were to be found). In fact, we show that families of orthonormal bases that form 
weighted complex projective 2-designs specify collections of orthogonal measurements which are (uniquely) optimal for quantum state 
tomography.

The article is organized as follows. In the next section we give a precise definition of a weighted $t$-design in $\C P^{d-1}$. 
In Sec.~\ref{sec:designbases} we translate this notion to the class of $t$-designs formed by the union of a 
family of orthonormal bases, proving existence in every dimension, and then revealing equivalence to complete sets of MUBs 
in the special case of 2-designs constructed from $d+1$ bases. In Sec.~\ref{sec:2designbases} we present new constructions of weighted 2-designs 
in terms of highly nonlinear functions on abelian groups. In fact, constructions using only $O(d^2)$ bases are shown to be sufficient, which can be reduced to 
$d+1$ whenever $d$ is a prime power (using a complete set of MUBs), or $kd+2$ whenever $kd+1$ is a prime power, for any positive integer $k$. 
We discuss weighted complex projective $2$-designs in the role of state determination in Sec.~\ref{sec:icpovm}, showing that such 
designs that are constructed from families of bases are optimal in this role in two specific scenarios: 
quantum state estimation as measurement-based cloning, and quantum state tomography by orthogonal measurements.
Finally, in Sec.~\ref{sec:conclude} we summarize our results.

\section{Weighted complex projective $t$-designs}
\label{sec:design}

The extension of spherical $t$-designs~\cite{Delsarte77} to projective spaces was first considered by 
Neumaier~\cite{Neumaier81}, but for the most part studied by Hoggar~\cite{Hoggar82,Hoggar84,Hoggar89,Hoggar92}, 
and, Bannai and Hoggar~\cite{Bannai85,Bannai89}. For a unified treatment of designs in terms of metric spaces 
consult the work of Levenshtein~\cite{Levenshtein92,Levenshtein98a,Levenshtein98b} 
(see also Ref.'s~\cite{Dunkl79,Seymour84,Nikova95,Boyvalenkov97,Nikova98,Boyvalenkov99,Konig99,Seidel01}). Our interest lies with the 
complex projective space $\C P^{d-1}$ of lines passing through the origin in $\C^d$. In this case each 
$x\in\C P^{d-1}$ may be represented by a unit vector $\ket{x}\in\C^d$ (modulo a phase), or more appropriately, 
by the rank-one projector $\pi(x)\coloneq\ketbra{x}$. We will use both representations in this article.
Roughly speaking, a complex projective $t$-design is then a finite subset of $\C P^{d-1}$ with the property that 
the discrete average of a polynomial of degree $t$ or less over the design equals the uniform average. Many
equivalent definitions can be made in these terms (see e.g. Ref.'s~\cite{Neumaier81,Hoggar82,Levenshtein92,Konig99,Seidel01}). 
In the general context of compact metric spaces, for example, Levenshtein~\cite{Levenshtein98a,Levenshtein98b} 
calls the pair $(\mathscr{D},w)$, where $\mathscr{D}$ a finite subset of $\C P^{d-1}$ and $w$ is positive-valued 
function on $\mathscr{D}$ with the normalization 
%\footnote{Levenshtein in fact chooses the weight function $m\coloneq|\mathscr{D}|\,w$.} 
$\sum_{x\in\mathscr{D}}w(x)=1$, a {\it weighted $t$-design\/} if 
\begin{equation}\label{eq:predesign}
\sum_{x,y\in\mathscr{D}}w(x)w(y)f\!\left(|\braket{x}{y}|^2\right)\;=\;
\iint_{\C P^{d-1}}\d\muu(x)\d\muu(y)\,f\!\left(|\braket{x}{y}|^2\right)
\end{equation}
for any real polynomial $f$ of degree $t$ or less, where $\muu$ denotes the unique unitarily invariant 
probability measure on $\C P^{d-1}$ induced by the Haar measure on $\mathrm{U}(d)$. When $w(x)=1/|\mathscr{D}|$
we recover the more common notion of an ``unweighted'' $t$-design. In the current context, however, 
it is appropriate to consider the more general notion, and then make an alternative explicit definition which is specialized to 
complex projective spaces. With this in mind, let $\Pisymt$ denote the projector onto the totally 
symmetric subspace of $(\C^d)^{\otimes t}$, which has dimension $\tbinom{d+t-1}{t}$, and recall that
\begin{equation}\label{eq:symavg}
\int_{\C P^{d-1}}\d\muu(x)\,\pi(x)^{\otimes t}\;=\;\tbinom{d+t-1}{t}^{-1}\,\Pisymt \;.
\end{equation}
Since the LHS is invariant under all unitaries $U^{\otimes t}$, which act irreducibly on the totally 
symmetric subspace of $(\C^d)^{\otimes t}$, Eq.~(\ref{eq:symavg}) follows from a straightforward application of Schur's Lemma.
The following definition of a weighted $t$-design is now equivalent to the one given above (we will defer the proof until the end of this section).

A countable set $\mathscr{S}$ endowed with a weight function $w:\mathscr{S}\rightarrow(0,1]$, normalized such that $\sum_{x\in\mathscr{S}}w(x)=1$, will be called a 
{\em weighted set\/} and denoted by the pair $(\mathscr{S},w)$. 
\begin{dfn}\label{dfn:wdesign}
A finite weighted set $(\mathscr{D},w)$, $\mathscr{D}\subset\C P^{d-1}$, is called a {\em weighted $t$-design (of dimension $d$)\/} if 
\begin{equation}\label{eq:wdesign}
\sum_{x\in\mathscr{D}}\,w(x)\pi(x)^{\otimes t} \;=\; \int_{\C P^{d-1}}\d\muu(x)\,\pi(x)^{\otimes t} \;=\; \tbinom{d+t-1}{t}^{-1}\,\Pisymt\;.
\end{equation}
\end{dfn}
 
Seymour and Zaslavsky have shown that (unweighted) $t$-designs in $\C P^{d-1}$ exist for any $t$ and $d$~\cite{Seymour84}. 
Notice that the normalization of $w$ is already implied by the trace of Eq.~(\ref{eq:wdesign}).
If we instead ``trace out'' only one subsystem of these $t$-partite operators, we can immediately deduce that every weighted 
$t$-design is also a weighted $(t-1)$-design. A weighted 1-design is known as a {\em tight frame\/} in the context of frame 
theory~\cite{Christensen03}, in which case the unnormalized states $\ket{\widetilde{x}}\coloneq\sqrt{w(x)d}\,\ket{x}$ 
are the frame vectors, and Eq.~(\ref{eq:wdesign}) is the tight frame condition: $\sum_{x\in\mathscr{D}}\ketbra{\widetilde{x}}=I$. 
In this form it is immediately apparent that we must have $|\mathscr{D}|\geq d$ for a weighted 1-design, with equality only 
if the frame vectors $\ket{\widetilde{x}}$ form an orthonormal basis for $\C^d$, i.e. $w(x)=1/d=1/|\mathscr{D}|$ and 
$|\braket{x}{y}|^2 = \delta(x,y)$ for all $x,y\in\mathscr{D}$. The 2-design case is treated in the following theorem
(see e.g. Ref.~\cite[Theorem 4]{Scott06} for a proof).

\begin{thm}\label{thm:2design}
Let $(\mathscr{D},w)$ be a weighted $2$-design of dimension $d$. Then $|\mathscr{D}|\geq d^2$ 
with equality only if $w(x)= 1/|\mathscr{D}|$ and 
\begin{equation}\label{eq:tight2designcondition}
|\braket{x}{y}|^2 \;=\; \frac{d\delta(x,y)+1}{d+1} \;,
\end{equation}
for all $x,y\in\mathscr{D}$. 
\end{thm}

Within the context of quantum information theory, a set of $d^2$ lines obeying Eq.~(\ref{eq:tight2designcondition}) is called a 
{\it symmetric IC-POVM (SIC-POVM)\/}~\cite{Renes04} (see also Ref.'s~\cite{Grassl04,Appleby05,Grassl05,Klappenecker05a,Colin05,Wootters06,Flammia06,Godsil05,Appleby06}). 
In general, the weighted complex projective 2-designs form a class of IC-POVMs which can be considered optimal in 
the role of state determination~\cite{Scott06}.
 
Theorem~\ref{thm:2design} is in fact a special case from known results within the theory of $t$-designs. In general, the number of 
design points must satisfy~\cite{Hoggar82,Bannai85,Dunkl79,Levenshtein92}
\begin{equation}\label{eq:designbound}
|\mathscr{D}|\;\geq\;\binom{d+\lceil t/2 \rceil -1}{\lceil t/2 \rceil}\binom{d+\lfloor t/2\rfloor -1}{\lfloor t/2\rfloor}\;,
\end{equation}
with equality only if the design has uniform weight~\cite{Levenshtein92}, i.e. $w(x)= 1/|\mathscr{D}|$.
A design which achieves this bound is called {\em tight\/}\footnote{The term ``tight'' is used differently in the contexts of frames and $t$-designs. 
A tight frame saturates the so-called frame bound [Eq.~(\ref{eq:welchbound}) with $t=1$] whereas a tight $t$-design saturates Eq.~(\ref{eq:designbound}).
Tight $t$-designs, being $1$-designs, are tight frames, but the converse need not be true.}. Tight $t$-designs in $\C P^1$ are equivalent to tight 
spherical $t$-designs on the Euclidean 2-sphere. Such designs exist only for $t=1,2,3,5$ (see e.g. Ref.~\cite{Hardin96}). 
When $d\geq 3$ it is known that tight $t$-designs in $\C P^{d-1}$ exist only for $t=1,2,3$~\cite{Bannai85,Bannai89,Hoggar89}. 
It is trivial that tight 1-designs exist in all dimensions. Tight 2-designs are conjectured to also exist in all 
dimensions~\cite{Zauner99,Renes04}. Analytical constructions, however, are known only for $d\leq 10$ and 
$d=12,13,19$~\cite{Zauner99,Renes04,Hoggar98,Grassl04,Appleby05,Grassl05}. Tight 3-designs may exist only in even dimensions.
Examples are known for $d=2,4,6$~\cite{Hoggar82}. Like in the specific 2-design case, more can be said about the 
structure of the tight $t$-designs. Given their rarity for higher values of $t$, however, we will defer further results in this direction 
to the work of Bannai and Hoggar~\cite{Hoggar82,Hoggar84,Hoggar89,Hoggar92,Bannai85,Bannai89}.
The task of finding $t$-designs is facilitated by the following theorem. 

\begin{thm}\label{thm:welch}
For any finite weighted set $(\mathscr{S},w)$, $\mathscr{S}\subset\C P^{d-1}$, and any $t\geq 1$,
\begin{equation}\label{eq:welchbound}
\sum_{x,y\in\mathscr{S}}w(x)w(y)\,|\braket{x}{y}|^{2t}\;\geq\;\tbinom{d+t-1}{t}^{-1}\;,
\end{equation}
with equality if and only if $(\mathscr{S},w)$ is a weighted $t$-design.
\end{thm}
\begin{proof}
Defining $S\coloneq\sum_{x\in\mathscr{S}}w(x)\pi(x)^{\otimes t}-\tbinom{d+t-1}{t}^{-1}\Pisymt$ we see that
\begin{equation}
0 \;\leq\; \tr(S^\dag S) \;=\; \sum_{x,y\in\mathscr{S}}w(x)w(y)\,|\braket{x}{y}|^{2t} - \tbinom{d+t-1}{t}^{-1}\;, 
\end{equation}
with equality if and only if $S=0$, which is the defining property of a $t$-design.
\end{proof}

This theorem allows us to check whether a weighted set of points in $\C P^{d-1}$ forms a $t$-design by considering only the 
angles between the supposed design elements. It also shows that $t$-designs can be found numerically by 
parametrizing a weighted set and minimizing the LHS of Eq.~(\ref{eq:welchbound}). The lower bound is in 
fact a straightforward generalization of the Welch bound~\cite{Welch74} (the above proof follows Ref.~\cite{Konig99}). 
We conclude this section by presenting two common alternative definitions of complex projective $t$-designs. The first 
was given at the outset [Eq.~(\ref{eq:predesign})].

\begin{prp}\label{prp:altdef1}
A finite weighted set $(\mathscr{D},w)$, $\mathscr{D}\subset\C P^{d-1}$, is a weighted $t$-design if and only if
\begin{equation}\label{eq:altdef1}
\sum_{x,y\in\mathscr{D}}w(x)w(y)f\!\left(|\braket{x}{y}|^2\right)\;=\;\iint_{\C P^{d-1}}\d\muu(x)\d\muu(y)\,f\!\left(|\braket{x}{y}|^2\right)
\end{equation}
for all polynomials $f$ of degree $t$ or less.
\end{prp}
\begin{proof}
Choosing the monomial $f(u)=u^t$ in Eq.~(\ref{eq:altdef1}) and integrating the RHS we obtain equality in Eq.~(\ref{eq:welchbound}). 
Thus by Theorem~\ref{thm:welch}, a finite weighted set $(\mathscr{D},w)$ satisfying Eq.~(\ref{eq:altdef1}) for $f(u)=u^t$ is a 
$t$-design. The converse is just as simple. By squaring both sides of Eq.~(\ref{eq:wdesign}) 
and then taking the trace, we see that Eq.~(\ref{eq:altdef1}) is satisfied by the monomial $f(u)=u^t$ when $(\mathscr{D},w)$ 
is a $t$-design. The same is true for any monomial of degree less than $t$, since a $t$-design is also a $(t-1)$-design, and thus 
by linearity, any polynomial of degree $t$ or less.
\end{proof}

Let $\{\ket{e_j}\}_{j=0}^{d-1}$ be the ``standard'' basis for $\C^d$. Define $\Hom(t,t)$ to be the set of polynomials which are 
homogeneous of degree $t$ in the coordinates $\braket{e_j}{x}={\rm x}_j$ on the unit sphere in $\C^d$, and also homogeneous of degree $t$ 
in the conjugates of these coordinates, $\braket{x}{e_j}=\overline{{\rm x}}_j$. For example, $f(x)=\braket{e_0}{x}\braket{x}{e_1}={\rm x}_0\overline{{\rm x}}_1$ is in $\Hom(1,1)$.
\begin{prp}\label{prp:altdef2}
A finite weighted set $(\mathscr{D},w)$, $\mathscr{D}\subset\C P^{d-1}$, is a weighted $t$-design if and only if
\begin{equation}\label{eq:altdef2}
\sum_{x\in\mathscr{D}}w(x)f(x)\;=\;\int_{\C P^{d-1}}\d\muu(x)\,f(x)
\end{equation}
for all polynomials $f\in\Hom(t,t)$.
\end{prp}
\begin{proof}
Simply note that Eq.~(\ref{eq:altdef2}) for each monomial $f\in\Hom(t,t)$ is given by a matrix component of 
Eq.~(\ref{eq:wdesign}) in the standard basis. The monomials form a basis for $\Hom(t,t)$.
\end{proof}

\section{Weighted $t$-designs from bases}
\label{sec:designbases}

We now address the problem of constructing weighted 2-designs in $\C P^{d-1}$ from 
the union of a family of orthonormal bases for $\C^d$. If the weight remains constant across 
elements of the same basis then such designs correspond to tight IC-POVMs~\cite{Scott06} 
which can be realized by a sequence of orthogonal measurements. We will thus make this a requirement.

To be precise, in the general case we seek a family of sets $\mathscr{B}_0,\dots,\mathscr{B}_{m-1}\subset\C P^{d-1}$, each  
specified by an orthonormal basis for $\C^d$, i.e. $\mathscr{B}_a=\{e^a_j\}_{j=0}^{d-1}$ where 
$\braket{e^a_j}{e^a_k}=\delta_{jk}$, and appointed a positive weight $w_a$, such that 
their union $\mathscr{D}=\cup_a\mathscr{B}_a$ forms a weighted $t$-design with the weight function 
$w(x)=\sum_a w_a 1_{\mathscr{B}_a}(x)$. The set indicator function, $1_A(x)\coloneq 1$ if $x\in A$ and 0 otherwise, 
takes care of any multiplicity in elements across different bases. Notice that the normalization of $w(x)$ implies normalization 
of the basis weights: $\sum_a w_a=1/d$. Although we in fact have $\mathscr{B}_a\subset\C P^{d-1}$, we will 
refer to $\mathscr{B}_a$ as an ``orthonormal basis for $\C^d$,'' and then the line $e^a_j\in\C P^{d-1}$ as 
the $j$-th element of the $a$-th basis. Revisiting Eq.~(\ref{eq:wdesign}) 
shows that we require
\begin{equation}\label{eq:wdesignbases}
\sum_{a=0}^{m-1}\,w_a\sum_{j=0}^{d-1}\,\pi(e^a_j)^{\otimes t} \;=\; \int_{\C P^{d-1}}\d\muu(x)\,\pi(x)^{\otimes t} \;=\; \tbinom{d+t-1}{t}^{-1}\,\Pisymt\;,
\end{equation}
or equivalently, by Theorem~\ref{thm:welch},
\begin{equation}\label{eq:welchbases}
\sum_{a,b=0}^{m-1}\, w_a w_b \sum_{j,k=0}^{d-1}\, |\braket{e^a_j}{e^b_k}|^{2t}\;=\;\tbinom{d+t-1}{t}^{-1}\;.
\end{equation}

Seymour and Zaslavsky~\cite{Seymour84} have given a non-constructive proof that 
(unweighted) \mbox{$t$-designs} exist in every dimension. Their main result is quite general, and also applies to 
weighted $t$-designs constructed from bases:

\begin{thm}\label{thm:SZ}
Let $\Om$ be a path-connected topological space endowed with a measure $\om$ that is finite and positive with 
full support, and, let $f: \Om \rightarrow \re^n$ be a continuous, integrable function. Then there 
exists a finite set $\mathscr{X}\subseteq\Om$ such that 
\begin{equation}\label{eq:SZ}
\frac{1}{|\mathscr{X}|}\sum_{x\in\mathscr{X}} f(x) \;=\; \frac{1}{\om(\Om)}\int_\Om\d\om(x)\, f(x) \;.
\end{equation}
The size of $\mathscr{X}$ may be any number, with a finite number of exceptions.
\end{thm}
\begin{cor}\label{cor:exist}
For each pair of positive integers $t$ and $d$, and for all sufficiently large $m$, there exist weighted 
$t$-designs for $\C P^{d-1}$ which are formed by taking the union of $m$ orthonormal bases for $\C^d$ (as described above).
\end{cor}
\begin{proof}
Let $\Om=\mathrm{U}(d)$ and $\om=\mu$, the Haar measure with $\mu(\mathrm{U}(d))=1$. Applying 
Theorem~\ref{thm:SZ} to 
\begin{equation}
f(U) \;\coloneq\; \frac{1}{d}\sum_j \left[U\pi(e_j)U^\dag\right]^{\otimes t}\;,
\end{equation}
which maps $\mathrm{U}(d)$ into $\End(\C^d)^{\otimes t}\cong\R^n$, where $n=2d^{2t}$ and $\{e_j\}_{j=0}^{d-1}$ is the 
standard basis, we know that for all sufficiently large $m$ there exist sets $\mathscr{X}=\{U_a\}_{a=0}^{m-1}\subset\mathrm{U}(d)$ 
with the property that [Eq.~(\ref{eq:SZ})]
\begin{align}
\frac{1}{md}\sum_{a,j}\left[U_a\pi(e_j){U_a}^\dag\right]^{\otimes t} &\;=\; \frac{1}{d} \sum_j\int_{\mathrm{U}(d)}\d\mu(U)\, \left[U\pi(e_j)U^\dag\right]^{\otimes t} \label{eq:existprf}\\
&\;=\; \tbinom{d+t-1}{t}^{-1}\,\Pisymt \;,
\end{align}
using Schur's Lemma for the integral. Now setting $\ket{e_j^a}=U_a\ket{e_j}$ we have our desired result, i.e. 
Eq.~(\ref{eq:wdesignbases}) with $w_a=1/md$.
\end{proof}

Our proof of Corollary~\ref{cor:exist} in fact shows that weighted $t$-designs formed by the unweighted 
union of orthonormal bases exist in all dimensions, i.e. we can take $w_a=1/md$. An example of a family of bases with this 
property is a {\em complete set of mutually unbiased bases\/}~\cite{Ivanovic81,Wootters89} 
(see also Ref.'s~\cite{Alltop80,Calderbank97,Bandyopadhyay02,Klappenecker04,Wocjan05,Planat06,Godsil05,Saniga04,Bengtsson05,Archer05,Wootters06,Aschbacher07,Boykin05,Grassl04,Bengtsson06,Butterley07,Klappenecker05b}). 
Two orthonormal bases, $\mathscr{B}_a=\{e^a_j\}_{j=0}^{d-1}$ and $\mathscr{B}_b=\{e^b_k\}_{k=0}^{d-1}$, are called 
{\em mutually unbiased\/} if 
\begin{equation}\label{eq:MUBsdef}
|\braket{e_j^a}{e_k^b}|^2 \;=\; \frac{1}{d} 
\end{equation}
for all $0\leq j,k\leq d-1$. A {\em complete\/} set of mutually unbiased bases (MUBs) is a set of $m=d+1$ orthonormal 
bases with the property that each pair are mutually unbiased. This is the maximum possible number.
Such sets are known to exist whenever $d$ is a prime power, but no other examples have been found. It is straightforward 
to confirm [via Eq.~(\ref{eq:welchbases})] that a complete set of MUBs will form a 2-design when $w_a=1/md$~\cite{Barnum00,Klappenecker05b}. 
Our next result (which can be regarded as the analogue of Theorem~\ref{thm:2design} for the current case) shows that such sets are optimal, in that 
we always need $m\geq d+1$ bases to construct a weighted 2-design, with equality only if the bases are mutually unbiased.

\begin{thm}\label{thm:MUBs}
Let $\mathscr{B}_0,\dots,\mathscr{B}_{m-1}\subset\C P^{d-1}$ be a family of orthonormal bases for $\C^d$ whose union 
$\mathscr{D}=\cup_a\mathscr{B}_a$ forms a weighted $2$-design with weight function $w(x)=\sum_a w_a 1_{\mathscr{B}_a}(x)$ 
for some choice of the positive constants $w_0,\dots,w_{m-1}$. Then $m\geq d+1$ with equality only if $w_a=1/md$ for all $a$ 
and the bases are pairwise mutually unbiased. 
\end{thm}
\begin{proof}
Theorem~\ref{thm:2design} with $|\mathscr{D}|=md$ immediately shows that we must have $m\geq d$, but since 
Eq.~(\ref{eq:tight2designcondition}) can not be satisfied by a family of orthonormal bases, we in fact need $m\geq d+1$. 
In the case of equality, note that by Theorem~\ref{thm:welch} we require [Eq.~(\ref{eq:welchbases}) with $t=2$]
\begin{equation}\label{eq:thmMUBsproofa}
d \sum_a w_a^2 + \sum_{a \neq b}w_aw_b \sum_{j,k} \lambda_{jk}^2 \;=\; \tbinom{d+1}{2}^{-1} \;,
\end{equation}
where we have defined the positive numbers $\lambda_{jk}\coloneq|\braket{e^a_j}{e^b_k}|^2$. Moreover, Theorem~\ref{thm:welch} 
implies that the LHS of Eq.~(\ref{eq:thmMUBsproofa}) is minimal with respect to the variables $w_a$ and $\lambda_{jk}$ 
under the appropriate constraints, two of which are $\sum_a w_a = 1/d$ and 
\begin{equation}\label{eq:thmMUBsproofb}
\sum_{j,k}\lambda_{jk}\;=\;\sum_{j,k}|\braket{e^a_j}{e^b_k}|^2\;=\;\tr(I\cdot I) \;=\; d \;.
\end{equation}
We will now minimize the LHS of Eq.~(\ref{eq:thmMUBsproofa}) under these two constraints. 
The minimum of $\sum_{j,k} \lambda_{jk}^2$ subject to Eq.~(\ref{eq:thmMUBsproofb}) occurs only when $\lambda_{jk}=1/d$ for all 
$0\leq j,k\leq d-1$, i.e., when $\mathscr{B}_a$ and $\mathscr{B}_b$ are mutually unbiased. 
Then the LHS of Eq.~(\ref{eq:thmMUBsproofa}) reduces to
\begin{equation}\label{eq:thmMUBsproofc}
d \sum_a w_a^2 + \sum_{a \neq b}w_aw_b \;=\; (d-1) \sum_a w_a^2 + \frac{1}{d^2} \;,
\end{equation}
and here the minimum (under $\sum_a w_a = 1/d$) occurs only when $w_a = 1/md$ for all $0 \leq a\leq m-1$.
With this value, Eq.~(\ref{eq:thmMUBsproofc}) reduces to the RHS of Eq.~(\ref{eq:thmMUBsproofa}) when $m=d+1$. Equality 
in Eq.~(\ref{eq:thmMUBsproofa}) thus requires the bases to be pairwise mutually unbiased and $w_a = 1/md$ whenever $m=d+1$. 
\end{proof}

In general, for each positive integer $t$ and $d$, we would like to know the quantity $M(t,d)$, which we use to denote the 
minimum number of orthonormal bases needed to construct a weighted $t$-design in $\C P^{d-1}$, or less ambitiously, bounds on this quantity. 
Trivially, $M(1,d)=M(t,1)=1$. Theorem~\ref{thm:MUBs} shows that $M(2,d)\geq d+1$, with equality when $d$ is a prime power. This 
theorem also shows that knowledge of $M(2,d)$ in general would solve the MUBs problem, i.e., imply the existence (or most likely, 
nonexistence) of complete sets of MUBs in dimensions which are not prime powers. As we remarked earlier, no such examples have 
been found. For the first exceptional dimension however, $d=6$, a tight $3$-design exists which is formed from the unweighted 
($w_a=1/md$) union of $m=21$ bases. We thus know that $7 \leq M(2,6)\leq M(3,6)=21$ (In fact $M(3,d)=d(d+1)/2$ when $d=2,4,6$, the 
dimensions of the only known tight 3-designs). However, in the next section we show that, quite remarkably, $M(2,6)\leq 8$, 
and moreover, $M(2,d)\leq d+2$ whenever $d+1$ is a prime power.

\section{Weighted $2$-designs from bases: constructions}
\label{sec:2designbases}

In this section we give analytic constructions of weighted $2$-designs using ``highly nonlinear functions'' on abelian groups. 
Nonlinear functions on finite fields have been studied extensively in the context of classical cryptography, and several 
authors~\cite{Nyberg94,Horadam03,Carlet04,Pott04} have extended those concepts to arbitrary finite abelian groups. Here we work 
with a class of functions called differentially $1$-uniform, so named for their resistance to differential 
cryptanalysis~\cite{Chabaud95}. 

Let $G$ and $H$ be abelian groups with $|G| \leq |H| < \infty$, let $f$ be a function from $G$ to $H$, and consider the number of 
solutions in $x$ to the equation
\begin{equation}
\label{eqn:lin}
f(x+a) - f(x) \;=\; b \;.
\end{equation}
If $(a,b) = (0,0)$ then Eq.~\eqref{eqn:lin} has $|G|$ solutions. There are also $|G|$ solutions whenever $f$ is linear or affine and 
$b = f(a)$. A function is therefore highly nonlinear if Eq.~\eqref{eqn:lin} has as few solutions as possible for any choice of $a$ 
and $b$. Of course we cannot avoid all solutions: letting $b = f(x+a)-f(x)$ for any fixed $x$ and $a$ gives at least one solution. 
The function $f$ is called {\em differentially $1$-uniform\/}, or simply {\em $1$-uniform\/}, if for every $(a,b) \neq (0,0)$, 
Eq.~\eqref{eqn:lin} has at most one solution~\cite{Nyberg94}. 

By way of example, let $\Z_n$ denote the cyclic abelian group of order $n$ and define $f: \Z_5 \rightarrow \Z_6$ through the following table.
\[
% \begin{array}{c|ccc}
% x & 0 & 1 & 2 \\
% \hline
% f(x) & 0 & 0 & 1
% \end{array} \qquad
\begin{array}{c|ccccc}
x & 0 & 1 & 2 & 3 & 4 \\
\hline
f(x) & 0 & 1 & 0 & 2 & 2 
\end{array}
\]
It is straightforward to verify that $f$ is differentially $1$-uniform. 

In constructing weighted $2$-designs we will need the following characterization.

% Carlet and Ding call a $f: G \rightarrow H$ a {\em perfect nonlinear} (PN) function if Eq.~\eqref{eqn:lin} has at most $|G|/|H|$ 
% solutions for $a \neq 0$. Implicit in their work is the assumption that $|G| \geq |H|$, as PN functions clearly do not exist for 
% $|G| < |H|$. When $|G| = |H|$, their PN functions and our $1$-uniform functions coincide. Pott calls $f:G \rightarrow H$ 
% {\em maximally nonlinear} (MN) if the maximum value of $|\sum_{x \in G} \chi(x)\psi(f(x))|$ over all nontrivial characters 
% $(\chi,\psi)$ of $G \times H$ is minimal. All PN functions are MN. 

\begin{lem}\label{lem:npnchar}
Let $G$ and $H$ be abelian groups. Then the function $f:G \rightarrow H$ is differentially $1$-uniform if and only if the equation
\begin{equation}\label{eqn:lin2}
(w,f(w)) + (x,f(x)) \;=\; (y,f(y)) + (z,f(z))
\end{equation}
has exactly $|G|(2|G|-1)$ solutions in $(w,x,y,z)$.
\end{lem}
\begin{proof}
Let $d=|G|$ and note that Eq.~\eqref{eqn:lin2} has $d(2d-1)$ trivial solutions, namely $d$ solutions of the form $(w,x,y,z) = (w,w,w,w)$ 
and $2(d^2-d)$ solutions of the form $(w,x,y,z) = (w,x,w,x)$ or $(w,x,y,z) = (w,x,x,w)$ for $x \neq w$. Now rewriting Eq.~\eqref{eqn:lin2} 
in the form
\begin{equation}
(w,f(w)) - (z,f(z)) \;=\; (y,f(y)) - (x,f(x)) \;=\; (a,b) \;,
\end{equation}
for some $a$ and $b$, we see that $f$ is $1$-uniform if and only if the solutions all satisfy $w = z$ or $w = y$. That is, $f$ is 
$1$-uniform if and only if the only solutions are the trivial ones.
\end{proof}

We now come to our main construction.

\begin{thm}\label{thm:npnbases}
Suppose there is a differentially $1$-uniform function $f:G \rightarrow H$ for the abelian groups $G$ and $H$. Then there exists a weighted 
$2$-design in $\C P^{|G|-1}$ which is formed from the union of $|H|+1$ orthonormal bases for $\C^{|G|}$.
\end{thm}

Setting $d=|G|$ and $m=|H|+1$, it follows that $M(2,d) \leq m$ whenever there exists a function $f:G \rightarrow H$ which is 
differentially $1$-uniform. The designs that Theorem~\ref{thm:npnbases} refers to are constructed as follows. We first assign the 
weight 
\begin{equation}\label{eq:weight1}
w_0 \;=\; \frac{1}{d(d+1)}
\end{equation}
to $\mathscr{B}_0 \coloneq \{e_j\}_{j=0}^{d-1}$, which is the standard basis for $\C^d$. All $m-1$ remaining bases are 
appointed the weight 
\begin{equation}\label{eq:weight2}
\qquad\qquad\quad   w_a \;=\; \frac{1}{(m-1)(d+1)} \qquad (a>0)\;,
\end{equation}
and then defined in terms of the characters of $G$ and $H$. For a review of characters of finite abelian groups, consult Ref.~\cite{Ledermann77}. 
Now for each $j \in G$, let $\chi_j$ be the $j$-th character of $G$, and similarly for each $a \in H$, let $\psi_a$ be the $a$-th character 
of $H$. The $j$-th element of basis $\mathscr{B}_a$ is then
\begin{equation}\label{eqn:basischar}
\qquad\qquad\quad \ket{e_j^a} \;\coloneq\; \frac{1}{\sqrt{d}} \sum_{x \in G} \chi_j(x)\psi_a\big(f(x)\big) \ket{e_x} \qquad (a>0)\;,
\end{equation}
using $0,\dots,d-1$ to denote the elements of $G$, but $1,\dots,m-1$ to denote the elements of $H$ (with $m-1$ now the 
additive identity), and the index $0$ always reserved for the standard basis in the latter context. The requirement  
$\braket{e_j^a}{e_k^a}=\delta_{jk}$ now follows from the orthogonality of characters.

\begin{proof}[Proof of Theorem~\ref{thm:npnbases}]
Let $d = |G|$ and $m=|H|+1$. By Theorem~\ref{thm:welch}, it suffices to show that 
\begin{equation}\label{eqn:npnsum}
\sum_{a,b \in H \cup \{0\}} \om_a \om_b \sum_{j,k \in G} \abs{\ip{e_j^a}{e_k^b}}^4 \;=\; \frac{2}{d(d+1)} 
\end{equation}
for the above weights [Eq.'s~\eqref{eq:weight1} and \eqref{eq:weight2}], and with $\mathscr{B}_a$ given by the standard basis
when $a=0$, i.e. $\ket{e^0_j} = \ket{e_j}$, or defined by Eq.~\eqref{eqn:basischar} otherwise.

When $a = b = 0$, we have $|\ip{e_j^0}{e_k^0}|^4 = |\ip{e_j}{e_k}|^4 =\delta_{jk}$. These terms contribute a total of
\begin{equation}\label{eq:sum1}
{\om_0}^2 \sum_{j,k \in G} \abs{\ip{e_j^0}{e_k^0}}^4 \;=\; \frac{1}{d^2(d+1)^2} \sum_{j,k \in G} \delta_{jk} \;=\; \frac{1}{d(d+1)^2} 
\end{equation}
to the LHS of Eq.~\eqref{eqn:npnsum}. When $a = 0$ and $b \in H$, we have 
$|\ip{e_j^0}{e_k^b}|^4 =|\ip{e_j}{e_k^b}|^4 = 1/d^2$, and likewise for $a \in H$ and $b=0$, adding a total of 
\begin{equation}\label{eq:sum2}
2\,\sum_{b \in H} \om_0 \om_b \sum_{j,k \in G} \abs{\ip{e_j^0}{e_k^b}}^4 \;=\; 2\,\sum_{b \in H} \frac{1}{d(m-1)(d+1)^2} \sum_{j,k \in G} \frac{1}{d^2} \;=\;\frac{2}{d(d+1)^2} 
\end{equation}
to the sum. 

For the remainder, we must evaluate $|\ip{e_j^a}{e_k^b}|^4$ for $a,b \in H$. First note that
\begin{align}
\ip{e_j^a}{e_k^b} & \;=\; \frac{1}{d} \sum_{x \in G}\overline{\chi_j(x)\psi_a\big(f(x)\big)} \chi_k(x)\psi_b\big(f(x)\big) \\
& \;=\; \frac{1}{d} \sum_{x \in G} \chi_{k-j}(x)\psi_{b-a}\big(f(x)\big) \;,
\end{align}
since the product of two characters, or the complex conjugate of a character, is another character. Multiplying by the conjugate, 
it follows that
\begin{align}
d^4 \abs{\ip{e_j^a}{e_k^b}}^4 & \;=\; \bigg(\sum_{x \in G} \chi_{k-j}(x)\psi_{b-a}\big(f(x)\big)\bigg)^2 \bigg(\sum_{y \in G}\overline{\chi_{k-j}(y)\psi_{b-a}\big(f(y)\big)}\bigg)^2 \\
& \;=\; \sum_{w,x,y,z \in G} \chi_{k-j}(w+x-y-z)\psi_{b-a}\big(f(w)+f(x)-f(y)-f(z)\big) \;.
\end{align}
Now taking the sum over $k \in G$ and $b \in H$, every character of $G$ and $H$ occurs once:
\begin{align}
d^4 \mathop{\sum_{k \in G}}_{b \in H} \abs{\ip{e_j^a}{e_k^b}}^4 & \;=\; \mathop{\sum_{k \in G}}_{b \in H} \; \sum_{w,x,y,z \in G} \chi_k(w+x-y-z)\psi_b\big(f(w)+f(x)-f(y)-f(z)\big) \qquad\\
& \;=\; \sum_{w,x,y,z \in G} \; \mathop{\sum_{k \in G}}_{b \in H} \chi_{w+x-y-z}(k)\psi_{f(w)+f(x)-f(y)-f(z)}(b) \label{eq:constructproof1}\\
& \;=\; d^2(m-1)(2d-1)\;. \label{eq:constructproof2} 
\end{align}
The last step is explained as follows. The inner summation in the RHS of Eq.~\eqref{eq:constructproof1} is zero unless both of the characters $\chi_{w+x-y-z}$ and $\psi_{f(w)+f(x)-f(y)-f(z)}$ are trivial, in which case 
the sum is $|G|\cdot|H|=d(m-1)$. But the characters are trivial exactly when $w+x =y+z$ and $f(w)+f(x) = f(y)+f(z)$. By Lemma~\ref{lem:npnchar}, this occurs exactly $d(2d-1)$ times if and only if $f$ is $1$-uniform. Thus Eq.~\eqref{eq:constructproof1} reduces to 
Eq.~\eqref{eq:constructproof2}. It follows that the total contribution to Eq.~\eqref{eqn:npnsum} of terms with $a,b \in H$ is 
\begin{align}
\sum_{a,b \in H} \om_a \om_b \sum_{j,k \in G} \abs{\ip{e_j^a}{e_k^b}}^4 &\;=\; \frac{1}{(m-1)^2(d+1)^2}\mathop{\sum_{j \in G}}_{a \in H}\mathop{\sum_{k \in G}}_{b \in H} \abs{\ip{e_j^a}{e_k^b}}^4   \\
&\;=\; \frac{1}{(m-1)^2(d+1)^2}\mathop{\sum_{j \in G}}_{a \in H}\frac{(m-1)(2d-1)}{d^2}  \\
&\;=\; \frac{2d-1}{d(d+1)^2} \;.\label{eq:sum3}
\end{align}

Finally, adding up all contributions [Eq.'s~\eqref{eq:sum1}, \eqref{eq:sum2} and \eqref{eq:sum3}], we find that Eq.~\eqref{eqn:npnsum} is satisfied, and so the union of the given bases forms a 
weighted $2$-design.
\end{proof}

We now turn to the problem of constructing differentially $1$-uniform functions. When $\nobreak{f:G \rightarrow H}$ is 
$1$-uniform, we always have $|G| \leq |H|$; in terms of minimizing the number of bases in a $2$-design, the goal is to minimize $|H|$. 
If $|H| = |G| = d$, then Theorem~\ref{thm:npnbases} produces a complete set of MUBs. Differentially $1$-uniform functions 
$f: G \rightarrow G$ are also called {\em perfect nonlinear}~\cite{Carlet04} or {\em maximally nonlinear}~\cite{Pott04}, and 
they are known to exist whenever $d$ is an odd prime power. In particular, let $\F_d$ be the finite field of order $d = p^n$. 
Then the function $f: \F_d \rightarrow \F_d$ is $1$-uniform in the following cases~\cite{Yuan06}:
\begin{enumerate}
\item $f(x) = x^2$\,;
\item $f(x) = x^{p^k + 1}$, \, $n/\gcd(n,k)$ odd\,;
\item $f(x) = x^{\frac{3^k + 1}{2}}$, \, $p = 3$, \, $k$ odd, \, $\gcd(n,k) = 1$\,;
%\item $f(x) = x^{\frac{3^k + 3}{2}}$, \, $p = 3$, \, $k$ even \cite{Helleseth97};
%\item $f(x) = x^{12}$, \, $p^n = 27$ \cite{Helleseth97};
\item $f(x) = x^{10}-ux^6-u^2x^2$, \, $p=3$, \, $n$ odd, \, $u \in \F_d^*$\,,
\end{enumerate}
where $\F_d^*$ is the multiplicative group of $\F_d$.
The first example reproduces the MUBs of Ivanovi\'c~\cite{Ivanovic81} and Wootters and Fields~\cite{Wootters89}.

When $d$ is an even prime power, $d = 2^n$ say, there are no $1$-uniform functions from $\F_d$ to $\F_d$. This is because in a field 
of characteristic $2$, solutions to the equation $f(x+a)-f(x)=b$ come in pairs $\{x,x+a\}$. However, there are $1$-uniform functions 
from $\F_d$ to $GR(4^n)$, the Galois ring of order $d^2$. For background on Galois rings, see Ref.'s~\cite{McDonald74} or \cite{Hammons94}. 
Let $\mathcal{T}$ be the Teichm\"uller set of $GR(4^n)$, and for each $x \in \F_d$ let $\hat{x}$ be the unique element of $\mathcal{T}$ 
such that $x \equiv \hat{x}$ mod $2$. Then $f(x) = \hat{x}$ is $1$-uniform~\cite{Hammons94}, and so there is a $2$-design formed from the 
union of $d^2+1$ bases for $\C^d$. In fact, all of the bases in this construction (except the standard basis) are repeated $d$ times, 
which amounts to giving each distinct basis a larger weight. The result is the complete set of $d+1$ MUBs described by Klappenecker 
and R{\"o}tteler~\cite{Klappenecker04}.

When $d$ is not a prime power, numerical evidence suggests that no complete set of MUBs exists. If this is true, then $M(2,d) \geq d+2$ 
for those values of $d$. The following construction shows that $M(2,d) \leq d+2$ whenever $d+1$ is a prime power.

\begin{prp}\label{prp:npn}
Let $d+1$ be a prime power, and let $y$ be a generator for $\F_{d+1}^*$. Then the function $f: \Z_d \rightarrow \F_{d+1}$ defined by
\begin{equation}
f(j) \; \coloneq \; y^j
\end{equation}
is differentially $1$-uniform.
\end{prp}

\begin{proof}
Suppose Eq.~\eqref{eqn:lin} has two solutions for some $a$ and $b$, say 
\begin{equation}
y^{j+a}-y^j \;=\; y^{k+a}-y^k \;.
\end{equation}
Factoring, we have
\begin{equation}
(y^j - y^k)(y^a - 1) \;=\; 0 \;,
\end{equation}
which implies that either $a = 0$ or $j = k$. So for $(a,b) \neq (0,0)$, Eq.~\eqref{eqn:lin} has at most one solution.
\end{proof}

% \begin{cor}
% If $d+1$ is a prime power, then $M(2,d) \leq d+2$.
% \end{cor}

The weighted $2$-designs resulting from Proposition~\ref{prp:npn} are explicitly constructed according to Eq.~\eqref{eqn:basischar} 
as follows. For $j\in\Z_d$, the $j$-th character of the group $\Z_d$ is $\chi_j(k) \coloneq e^{2\pi i jk/d}$; for $a \in \F_{d+1}$, 
the $a$-th character of $\F_{d+1}$ is $\psi_a(x) \coloneq  e^{2\pi i\tr(ax)/p}$, where $d+1=p^n$ ($p$ prime) and 
$\tr x\coloneq x+x^p+\dots +x^{p^{n-1}}$ is the trace function from $\F_{d+1}$ to $\F_p$. Thus if $y$ is a primitive element of 
$\F_{d+1}$, then the $j$-th vector of basis $\mathscr{B}_a$ is 
\begin{equation}
\ket{e_j^a} \;\coloneq\; \frac{1}{\sqrt{d}} \sum_{k=0}^{d-1} e^{2\pi i jk/d}e^{2\pi i\tr(ay^k)/p}\ket{e_k} \;.
\end{equation}

It remains to consider upper bounds on $M(2,d)$ when neither $d$ nor $d+1$ is a prime power. In these cases, the following proposition 
shows that $M(2,d) \leq kd+2$, where $k$ is the smallest positive number such that $kd+1$ is a prime power. The proof of 
Proposition~\ref{prp:npngen} is the same as that of Proposition~\ref{prp:npn}.

\begin{prp}\label{prp:npngen}
Let $kd+1$ be a prime power, and let $y$ be an element of multiplicative order $d$ in $\F_{kd+1}$. Then $f(j) \coloneq y^j$ is a 
differentially $1$-uniform function from $\Z_d$ to $\F_{kd+1}$.
\end{prp}

The following table summarizes the resulting best known upper bound on $M(2,d)$ for dimension $d \leq 50$: 
\[
\begin{array}{c|ccccccccccccc}
d & p^n & p^n-1 & 14 & 20 & 21 & 33 & 34 & 35 & 38 & 39 & 44 & 45 & 50 \\
\hline
M(2,d) \leq & d+1 & d+2 & 30 & 42 & 44 & 68 & 104 & 72 & 192 & 80 & 90 & 182 & 102
\end{array}
\]

A numerical search in dimension $d=14$ was unable to locate an example of a weighted 2-design composed of $m<30$ 
orthonormal bases. The method used, however, which is an optimization procedure based on Theorem~\ref{thm:welch}, is quite slow 
for such a large value of $d$ and should not be trusted.

In general, the upper limit for $M(2,d)$ obtained from Proposition \ref{prp:npngen} is far from the lower bound of $d+1$. 
Linnik's theorem~\cite{Linnik44} gives an upper bound on the size of the smallest prime that occurs in the arithmetic progression 
$(kd+1)_{k=1}^{\infty}$: it implies that $M(2,d)$ is $O(d^L)$, for some constant $L$. Heath-Brown~\cite{Heath-Brown92} has shown 
that $L \leq 5.5$. However, the next construction shows that $M(2,d)$ is $O(d^2)$.

\begin{prp}\label{prp:quadfunc}
Let $f:\Z_d \rightarrow \Z_n$ be the function $f(j) \coloneq \tbinom{j}{2}$, for $0 \leq j \leq d-1$. If $d>2$ and
$n \geq \frac{3}{4}(d-1)^2$, then $f$ is differentially $1$-uniform.
\end{prp}

\begin{proof}
We work over the integers: let $\hat{f}: \Z_d \rightarrow \Z$ be the function that maps $j$ to $\tbinom{j}{2}$, for $0 \leq j \leq d-1$. It then suffices to show that for every $a \in \{1,\ldots,d-1\}$ and $b \in \{0,\ldots,n-1\}$, the equation 
\begin{equation}
\hat{f}(j+a)-\hat{f}(j) \;\equiv\; b \mod n
\end{equation}
has at most one solution.  Fix $a \in \{1,\ldots,d-1\}$, and consider the differences
\begin{equation}
\hat{f}(j+a)-\hat{f}(j) \;=\; \begin{cases}
\tbinom{j+a}{2}-\tbinom{j}{2}\,, & j+a \leq d-1\,; \\
\tbinom{j+a-d}{2} - \tbinom{j}{2}\,, & j+a \geq d\,.
              \end{cases}
\end{equation}
First examine the cases in which $j+a \leq d-1$. For these values of $j$, the differences $\tbinom{j+a}{2}-\tbinom{j}{2}$ are all distinct mod $n$. 
For, if $\tbinom{j+a}{2}-\tbinom{j}{2} \equiv \tbinom{k+a}{2}-\tbinom{k}{2}$ mod $n$, then simplifying we find that 
\begin{equation}
a(j-k) \;\equiv\; 0 \mod n,
\end{equation}
which holds only if $j = k$, since $|a(j-k)|\leq a(d-1-a)\leq \frac{1}{4}(d-1)^2<n$ whenever $n \geq \frac{3}{4}(d-1)^2$ and 
$0\leq j,k \leq d-1-a$. Similarly, the differences for which $j+a \geq d$ are also distinct mod $n$ for distinct values of $j$. 

We now show that no difference with $j+a \leq d-1$ has the same value mod $n$ as a difference with $j+a \geq d$. The largest value of the former 
occurs when $j+a = d-1$, and the smallest of the latter occurs when $j = d-1$. So we require
\begin{equation}
n \;>\; \Big[\tbinom{d-1}{2}-\tbinom{d-a-1}{2}\Big] - \Big[\tbinom{a-1}{2} - \tbinom{d-1}{2}\Big]\;,
\end{equation}
which simplifies to 
\begin{equation}\label{eqn:quaddiff}
n \;>\; da - a^2 + \frac{d^2 - 3d}{2}\;.
\end{equation}
The largest value of the RHS of Eq.~\eqref{eqn:quaddiff} occurs at $a = \lfloor d/2 \rfloor$, which means the inequality is satisfied 
for all $a$ whenever $n \geq \frac{3}{4}(d-1)^2$. Finally, the smallest difference with $j+a \leq d-1$ needs to be 
greater than the largest difference with $j+a \geq d$, i.e.  $\tbinom{a}{2}-\tbinom{0}{2}>\tbinom{0}{2} - \tbinom{d-a}{2}$,
which is satisfied for all $a$ whenever $d>2$.
\end{proof}

It is also of some mathematical interest to construct weighted $2$-designs in $\C P^{d-1}$ which are the union of $m$ orthonormal bases for $m$ larger than the minimum $M(2,d)$. While these constructions are not so important in the context of quantum state tomography, they may be of use in other applications of designs. Moreover, insight into the general structure of such designs may eventually lead to improved bounds for $M(2,d)$. Numerically, it becomes easier to find designs with $m$ bases as $m$ increases, and 
Corollary~\ref{cor:exist} and Proposition~\ref{prp:quadfunc} indicate that such designs will always exist for sufficiently large $m$. 
We now use differentially $1$-uniform functions to more precisely quantify ``sufficiently large $m$'' for certain values of $d$.

There is very little literature on highly nonlinear functions $f: G \rightarrow H$ with $|G| < |H|$, as cryptographers have focused on 
the case $|G| \geq |H|$. Nevertheless, $1$-uniform functions become easier to find as $|H|$ increases. In fact, for fixed $G$, a random 
function $f: G \rightarrow H$ is asymptotically almost surely $1$-uniform as $|H| \rightarrow  \infty$. There are also many recursive 
constructions. For example: if $f_1:G \rightarrow H_1$ is any function and $f_2: G \rightarrow H_2$ is $1$-uniform, then the function 
$f_1+f_2: G \rightarrow H_1 \times H_2$ defined by 
\begin{equation}
(f_1+f_2)(x) \;\coloneq\; (f_1(x),f_2(x))
\end{equation}
is also $1$-uniform. Embedding the codomain of a $1$-uniform function into a larger group serves as a general strategy for constructing 
designs with many bases.

\begin{lem}\label{lem:npnembed}
Let $E:\Z_k \rightarrow \Z_n$ be the function that maps $(i \mod k)$ to $(i \mod n)$, for $0 \leq i \leq k-1$. If $f: G \rightarrow \Z_k$ 
is differentially $1$-uniform and $n \geq 2k-1$, then $E \circ f: G \rightarrow \Z_n$ is also differentially $1$-uniform. 
\end{lem}

\begin{proof}
We again work over the integers. Let $\hat{f}: G \rightarrow \Z$ be the function that maps $x$ to the unique integer 
$\hat{f}(x) \in \{0,\ldots,k-1\}$ such that $f(x) \equiv \hat{f}(x) \mod k$. Since $f$ is $1$-uniform, it follows that for fixed $a \neq 0$ 
and $b \in \Z$,
\begin{equation}
\hat{f}(x+a)-\hat{f}(x) \;=\; b
\end{equation}
has at most one solution. Moreover, since $\hat{f}(x)$ is in the range $[0,k-1]$, it is clear that $\hat{f}(x+a)-\hat{f}(x)$ is in 
$[-k+1,k-1]$, a range of size $2k-2$. But $n > 2k-2$, so it follows that $\hat{f}(x+a)-\hat{f}(x) = b$ has at most one solution mod $n$. 
Thus $E \circ f$ is $1$-uniform as a function from $G$ to $\Z_n$.
\end{proof}

The proof of Lemma~\ref{lem:npnembed} also demonstrates that any cyclic subgroup in the codomain of a $1$-uniform function can be embedded 
into a larger group. More precisely, suppose that $f_1: G \rightarrow H_1$ and $f_2:G \rightarrow \Z_k$ are functions such that $f_1 + f_2$ 
is $1$-uniform from $G$ to $H_1 \times \Z_k$. Then for any $n \geq 2k-1$, the function $f_1 + (E \circ f_2)$ is $1$-uniform from $G$ to  
$H_1 \times \Z_n$. But every group has some cyclic subgroup, so: if $f: G \rightarrow H$ is $1$-uniform, then for every $n \geq 2|H| - 1$, 
there a $1$-uniform function $g: G \rightarrow H'$ such that $H'$ is a group of order $n$.

% More generally, any cyclic subgroup in the codomain of a $1$-uniform function can be embedded into a larger group. The proof of 
% Lemma~\ref{lem:npnsubgroup} is the similar to that of Lemma~\ref{lem:npnembed}. 
%
% \begin{lem}
% \label{lem:npnsubgroup}
% Let $I_H$ be the identity function on group $H$. If $f$ is a differentially $1$-uniform function from $G$ to $H \times \Z_k$, then for 
% any $n \geq 2k-1$, $(I_H \times E) \circ f$ is a differentially $1$-uniform function from $G$ to $H \times \Z_n$.
% \end{lem}
%
% Lemma \ref{lem:npnsubgroup} implies that if $f: G \rightarrow H_1$ is $1$-uniform, then for every $n \geq 2|H_1| - 1$, there a 
% $1$-uniform function $g: G \rightarrow H_2$ where $H_2$ is a group of order $n$.

\begin{cor}
If $f: G \rightarrow H$ is differentially $1$-uniform, then for every $m \geq 2|H|$ there is weighted $2$-design in $\C P^{|G|-1}$ which 
is formed from the union of $m$ orthonormal bases for $\C^{|G|}$.
\end{cor}

Considering cyclic subgroups $\Z_p$ in the codomains of perfect nonlinear functions or the $1$-uniform functions in 
Proposition~\ref{prp:npn} yields the following.

\begin{cor}
Suppose $d = p^n$ with $p$ an odd prime, or $d + 1 = p^n$ with $p$ any prime. Then for every $m \geq d+p+1$, there is a weighted 
$2$-design in $\C P^{d-1}$ which is formed from the union of $m$ orthonormal bases for $\C^d$.
\end{cor}

\section{Weighted 2-designs as informationally complete POVMs}
\label{sec:icpovm}

The outcome statistics of a quantum measurement are described by a positive-operator-valued measure (POVM)~\cite{Busch96}
on a set $\mathscr{X}$ of measurement outcomes. When $\mathscr{X}$ is countable the POVM is completely characterized by a 
set of positive operators, $\{F(x)\}_{x\in\mathscr{X}}$, called the ``POVM elements,'' which together satisfy the normalization 
constraint $\sum_{x\in\mathscr{X}} F(x) = I$. An {\em informationally complete POVM (IC-POVM)\/}~\cite{Renes04,Scott06,Prugovecki77,Schroeck89,Busch89,Busch91} 
is one with the property that for each quantum state, $\rho\in\Qd\coloneq \big\{A\in\End(\C^d)\,|\,A\geq 0\,,\,\tr(A)=1\big\}$, 
the outcome statistics, $p(x)\coloneq\tr[F(x)\rho]$, uniquely identify the state. A sequence of measurements on copies of a 
system in an unknown state, enabling an estimate of the statistics, will then reveal the state. 

In this article we are dealing primarily with {\em rank-one\/} POVMs. It is then appropriate to consider the measurement 
outcomes as points in complex projective space, $\mathscr{X}\subseteq\C P^{d-1}$, and set $F(x)=\tau(x)\pi(x)$, where 
$\pi(x)\coloneq\ketbra{x}$ and the positive weights $\tau(x)$ inherit the normalization $\sum_{x\in\mathscr{X}}\tau(x)=d$. 
Important examples of such IC-POVMs include symmetric IC-POVMs (SIC-POVMs)~\cite{Renes04} and complete sets of mutually 
unbiased bases (MUBs)~\cite{Ivanovic81,Wootters89}. The main purpose of this section is to show that weighted 
complex projective 2-designs of the type constructed in the previous section, which include complete sets of MUBs as  
examples, specify optimal IC-POVMs for quantum state tomography by orthogonal measurements. This will be done in 
Sec.~\ref{subsec:qst}. We will begin by revisiting some of the results of Ref.~\cite{Scott06}.

It is clear that weighted $1$-designs are equivalent to rank-one POVMs under the association $\tau(x)=w(x)d$ and 
$\mathscr{X}=\mathscr{D}$. Weighted $2$-designs have the additional property of being informationally complete. A productive way 
of showing this is as follows. Equipped with the Hilbert-Schmidt inner product $\BraKet{A}{B}\coloneq\tr(A^\dag B)$, the vector 
space $\End(\C^d)\cong\C^{d^2}$ is an inner product space where we think of $\Bra{A}$ as an operator ``bra'' and $\Ket{B}$ as an 
operator ``ket'' (see Caves~\cite{Caves99} or Ref.~\cite{Scott06} for notational clarification). Addition and scalar 
multiplication of operator kets then follows that for operators, e.g. $\Ket{aA+bB}=a\Ket{A}+b\Ket{B}$ for $a,b\in\C$. Under 
the identification $A\otimes B^\dag\leftrightarrow\Ket{A}\Bra{B}$ we can rewrite our definition of a weighted 2-design 
[Eq.~(\ref{eq:wdesign}) with $t=2$ and $w(x)=\tau(x)/d$],
\begin{equation}\label{eq:superisom1}
\sum_{x\in\mathscr{D}}\,\tau(x)\pi(x)\otimes\pi(x) \;=\; \frac{2\Pisym^{(2)}}{d+1} \;=\; \frac{1}{d+1}\bigg(\sum_{j,k}\ket{e_j}\bra{e_k}\otimes\ket{e_k}\bra{e_j}\, +\, I\otimes I \bigg)\;,
\end{equation}
in superoperator notation as
\begin{equation}\label{eq:superisom2}
\sum_{x\in\mathscr{D}}\,\tau(x)\KetBrab{\pi(x)} \;=\;\frac{\Is+\KetBra{I}}{d+1} \;=\; \frac{1}{d+1}\bigg(\sum_{j,k}\KetBrab{\ket{e_j}\bra{e_k}}\, +\, \KetBra{I}\bigg)\;,
\end{equation}
where $\Is\coloneq\sum_{j,k}\KetBrab{\ket{e_j}\bra{e_k}}$ is the identity superoperator under the ``left-right'' 
action~\cite{Caves99} (meaning superoperators act on operators just like operators on vectors), i.e. 
$\Is\Ket{A}=\Ket{A}$ for all $A\in\End(\C^d)$. The informational completeness of $\{F(x)=\tau(x)\pi(x)\}_{x\in\mathscr{D}}$ 
is now immediately apparent from Eq.~(\ref{eq:superisom2}). In fact, an explicit state-reconstruction formula follows from the 
left-right action of this equation on a quantum state:
\begin{equation}
\sum_{x\in\mathscr{D}}\,\tau(x)\Ketb{\pi(x)}\BraKetb{\pi(x)}{\,\rho\,} \;=\;\frac{\Is\Ket{\rho}+\Ket{I}\BraKet{I}{\rho}}{d+1}\;=\;\frac{\Ket{\rho}+\Ket{I}}{d+1}
\end{equation}
which simplifies to
\begin{equation}\label{eq:reconstruct}
\rho \;=\; (d+1)\sum_{x\in\mathscr{D}}\,p(x)\pi(x) \;-\; I \;,
\end{equation}
where $p(x)\coloneq\tr[F(x)\rho]=\tau(x)\tr[\pi(x)\rho]$ are the measurement outcome statistics.

The map $\pi$ embeds complex projective space into $\End(\C^d)$. If we instead embed $\C P^{d-1}$ into the real vector space of traceless 
Hermitian operators $\Hd\coloneq\{A\in\End(\C^d)\,|\,A^\dag=A\,,\,\tr(A)=0\}\cong\R^{d^2-1}$, via the mapping 
$x\rightarrow\vartheta(x)\coloneq\pi(x)-I/d$, then Eq.~(\ref{eq:superisom2}) takes a revealing form.
Recalling the normalization $\sum_{x\in\mathscr{D}} \tau(x)=d$, we can rewrite this equation as
\begin{equation}\label{eq:superisom3}
\sum_{x\in\mathscr{D}}\,\tau(x)\KetBrab{\pi(x)-I/d} \;=\; \frac{\Is-\KetBra{I}/d}{d+1} \;=\; \frac{\Ptr}{d+1}\;,
\end{equation}
where $\Ptr\coloneq\Is-\KetBra{I}/d$ is the projector onto the subspace of traceless operators in $\End(\C^d)$. In $\Hd$, this 
of course means
\begin{equation}\label{eq:superisom4}
\sum_{x\in\mathscr{D}}\,\tau(x)\KetBrab{\vartheta(x)} \;=\; \frac{\Ih}{d+1}\;,
\end{equation}
where $\Ih$ denotes the identity superoperator for this space. The interpretation afforded by Eq.~(\ref{eq:superisom4}), 
which states that $\{\vartheta(x)\}_{x\in\mathscr{D}}$ forms a {\em tight (operator) frame\/}~\cite{Christensen03} in $\Hd$ 
with respect to the ``trace'' measure $\tau$, is that rank-one IC-POVMs which correspond to weighted 2-designs in $\C P^{d-1}$ are ``as close as possible''~\cite{Daubechies86,Casazza06} to orthonormal bases for $\Hd$, when embedded into this space. Weighted 
complex projective 2-designs have thus been called {\em tight\/} rank-one IC-POVMs~\cite{Scott06}. This analogy with tight 
frames is particularly pleasing since under the projection $\Ket{\rho}\rightarrow\Ptr\Ket{\rho}=\Ket{\rho-I/d}$, $\Hd$ is the 
natural place to study a general quantum state $\rho\in\Qd$. Indeed, $\Q2$ then corresponds to the Bloch sphere.

\subsection{Optimal quantum state estimation}
\label{subsec:qse}

Not only do tight rank-one IC-POVMs possess the above elegant structure, they are the optimal choice in the following 
state-estimation scenario. Consider a measuring instrument in the role of a cloning 
machine~\cite{Massar95,Gisin97,Derka98,Latorre98,Bruss99} (i.e. a one-to-infinity cloner \cite{Bae06}). 
The input to this machine is a single copy of an unknown pure state, $\psi$, and the output, $\hat{\rho}(x)$, is 
a state chosen to estimate $\psi$ based on a measurement (with outcome $x$). The fidelity between the input and output states,
averaged over the measurement outcomes, 
\begin{equation}
f^{(F,\hat{\rho})}(\psi) \;\coloneq\; \sum_{x\in\mathscr{X}} \tr\!\big[F(x)\pi(\psi)\big]\tr\!\big[\hat{\rho}(x)\pi(\psi)\big] \;,
\end{equation}
can then be optimized for a worst-case input, only when the measurement is described by a tight IC-POVM~(see Ref.'s~\cite{Scott06,Hayashi05}). 
More precisely,
\begin{equation}\label{eq:cloning}
f_\mathrm{wc}^{(F,\hat{\rho})} \;\coloneq\; \inf_{\psi\in\C P^{d-1}}\; f^{(F,\hat{\rho})}(\psi) \;\leq\; \frac{2}{d+1} \;,
\end{equation}
with equality if and only if the outcome statistics for the measurement are described by a tight rank-one IC-POVM, 
$\{F(x)=\tau(x)\pi(x)\}_{x\in\mathscr{X}}$ where $\mathscr{X}\subseteq\C P^{d-1}$, and $\hat{\rho}(x)=\pi(x)$ is 
chosen for the output state. In such cases the output fidelity is in fact independent of the input state: 
$f^{(F,\hat{\rho})}(\psi)=2/(d+1)$.

A measurement in a random basis is one strategy to achieve equality in Eq.~(\ref{eq:cloning}). However a measuring 
instrument configurable to only $m$ different bases, $\mathscr{B}_0,\dots,\mathscr{B}_{m-1}$, will also suffice, provided 
$\mathscr{D}=\cup_a\mathscr{B}_a$ forms a weighted $2$-design with weight $w(x)=\sum_a w_a 1_{\mathscr{B}_a}(x)$, and the 
\mbox{$a$-th} basis is chosen with probability $v_a = w_ad$. This follows from the straightforward fact that rolling an $m$-sided 
die, where the $a$-th side occurs with probability $v_a$, and then performing an orthogonal measurement in a basis 
corresponding to the result, $\mathscr{B}_b$ say, is one way of realizing the POVM 
$\{F(x)=\tau(x)\pi(x)\}_{x\in\mathscr{X}}$ with $\mathscr{X}=\cup_a\mathscr{B}_a$ and $\tau(x)=\sum_a v_a 1_{\mathscr{B}_a}(x)$. When $d$ is prime power we know that only $m=d+1$ configurations for the orthogonal measurements are needed, each specified by a member of a complete set of 
MUBs. If $d$ is not a prime power, but $d+1$ is, then the constructions of weighted 2-designs in the previous section 
show that $m=d+2$ configurations suffice. In general, $M(2,d)$ configurations are sufficient and necessary under 
a restriction to orthogonal measurements.

\subsection{Optimal quantum state tomography by a repeated general measurement}
\label{subsec:qstg}

Tight rank-one IC-POVMs are also an outstanding choice for {\em quantum state tomography\/}~\cite{Paris04}. Suppose that we 
are instead given $N$ copies of a system in an unknown general quantum state $\rho\in\Qd$. A sequence of measurements on these 
copies, each with a measuring instrument described by the same IC-POVM, $\{F(x)\}_{x\in\mathscr{X}}$ say, will 
provide an estimate of the statistics $p(x)=\tr[F(x)\rho]$, and hence, identify the state. 

State reconstruction for a general IC-POVM is facilitated by a {\em dual frame\/}~\cite{Christensen03} to the frame of POVM elements 
$\{F(x)\}_{x\in\mathscr{X}}$, i.e. a set of operators $\{Q(x)\}_{x\in\mathscr{X}}\subseteq\End(\C^d)$ satisfying
\begin{equation}\label{eq:dualframe}
\sum_{x\in\mathscr{X}} \Ketb{Q(x)}\Brab{F(x)} \;=\;\sum_{x\in\mathscr{X}}\tau(x) \Ketb{Q(x)}\Brab{P(x)} \;=\;  \Is\;,
\end{equation}
where we have introduced the positive-operator-valued density (POVD) $P(x)\coloneq F(x)/\tau(x)$ and $\tau(x)\coloneq \tr[F(x)]$ for a general POVM.
Alternatively, $Q$ is a dual frame to $P$ with respect to the trace measure $\tau$.
In the current context we will refer to $Q$ as a {\em reconstruction operator-valued density (OVD)} for the IC-POVM $F$.
The left-right action of the dual frame condition [Eq.~(\ref{eq:dualframe})] on a quantum state $\Ket{\rho}$ provides a state-reconstruction 
formula:
\begin{equation}\label{eq:linearreconstruct}
\rho \;=\; \sum_{x\in\mathscr{X}} p(x)Q(x)\;.
\end{equation}
There are generally many different choices for $Q$. The {\em canonical dual frame\/} to $P$ (with respect to $\tau$),
\begin{equation}\label{eq:canonicaldual}
\Ketb{R(x)} \;\coloneq\; \mathcal{F}^{-1}\Ketb{P(x)}\;,
\end{equation}
is found through the (left-right) inverse of the POVM superoperator,
\begin{equation}
\mathcal{F} \;\coloneq\; \sum_{x\in\mathscr{X}} \tau(x)\KetBrab{P(x)}\;.
\end{equation}
It is straightforward to confirm that $\mathcal{F}^{-1}$ exists if and only if the POVM is informationally complete, 
and that Eq.~(\ref{eq:dualframe}) is satisfied for $Q=R$. The optimality of this choice  
was established in Ref.~\cite{Scott06} for the current setting and then again by D'Ariano and Perinotti~\cite{DAriano07} 
for a similar scenario. In the special case of a tight rank-one IC-POVM, in which case $F(x)=\tau(x)\pi(x)$ and 
$\mathscr{X}\subseteq\C P^{d-1}$, the canonical dual is $R(x)=(d+1)\pi(x)-I$, and Eq.~(\ref{eq:linearreconstruct}) reduces to
Eq.~(\ref{eq:reconstruct}).

Now returning to our problem of state reconstruction, if $y_1,\dots,y_N\in\mathscr{X}$ are the measurement results, 
then one estimate for the statistics is simply
\begin{equation}\label{eq:pestimator}
\hat{p}(x)\;=\;\hat{p}(x;y_1,\dots,y_N)\;\coloneq\;\frac{1}{N}\sum_{k=1}^N\delta(x,y_k)\;, 
\end{equation}
which under Eq.~(\ref{eq:linearreconstruct}) gives
\begin{equation}\label{eq:linearreconstructe}
\hat{\rho} \;=\; \hat{\rho}(y_1,\dots,y_N) \;\coloneq\; \sum_{x\in\mathscr{X}} \hat{p}(x;y_1,\dots,y_N)Q(x)\;,
\end{equation}
for an estimate of $\rho$. We will call $\hat{\rho}$ a {\em linear tomographic estimate\/} of $\rho$ to distinguish 
it from more sophisticated choices, such as those from maximum likelihood estimation~\cite{Hradil97,Banaszek99} or Bayesian mean 
estimation~\cite{Jones91,Buzek98,Schack01,Tanaka05,BlumeKohout06}. 

The Hilbert-Schmidt distance $d_{\textrm{HS}}(\rho,\hat{\rho})\coloneq\|\rho-\hat{\rho}\|$, where 
$\|A\|\coloneq\sqrt{\BraKet{A}{A}}$, provides a measure of the expected error in our estimate:
\begin{equation}\label{eq:error}
e^{(F,Q)}(\rho) \;\coloneq\; \mathop{\textrm{\large E}}_{y_1,\dots,y_N}\Big[\,\|\rho-\hat{\rho}(y_1,\dots,y_N)\|^2\,\Big]\;.
\end{equation}
Let $\rho=\rho(\sigma,U)\coloneq U\sigma U^\dag$ for some fixed quantum state $\sigma\in\Qd$. It has been shown that, for random 
Hilbert-space orientations $U$ between the state $\sigma$ and measuring apparatus, the average Hilbert-Schmidt error can be minimized only when 
$\{F(x)\}_{x\in\mathscr{X}}$ is a tight IC-POVM (see Ref.~\cite[Theorem~18]{Scott06}). That is,
\begin{equation}\label{eq:optimalstatetomography}
e_\mathrm{av}^{(F,Q)}(\sigma) \;\coloneq\; \int_{\mathrm{U}(d)}\d\muu(U)\: e^{(F,Q)}(\rho(\sigma,U)) \;\geq\; \frac{1}{N}\Big(d(d+1)-1-\tr(\sigma^2)\Big) \;,
\end{equation}
with equality if and only if the outcome statistics for the measurements are described by a tight rank-one IC-POVM, 
$\{F(x)=\tau(x)\pi(x)\}_{x\in\mathscr{X}}$ where $\mathscr{X}\subseteq\C P^{d-1}$, and $Q(x)=R(x)=(d+1)\pi(x)-I$ is chosen for the dual frame. The same is true for the worst-case error 
$e_\mathrm{wc}^{(F,Q)}(\sigma) \coloneq \sup_{U\in\mathrm{U}(d)}e^{(F,Q)}(\rho(\sigma,U))$, and in fact, tight rank-one IC-POVMs 
form the unique class of POVMs achieving
\begin{equation}
e_\mathrm{wc}^{(F,R)}(\sigma) \;=\; e_\mathrm{av}^{(F,R)}(\sigma) \;=\; e^{(F,R)}(\rho(\sigma,U))  \;=\; \frac{1}{N}\Big(d(d+1)-1-\tr(\sigma^2)\Big) \;.
\end{equation} 

These results show that if we treat a family of $m$ bases $\mathscr{B}_0,\dots,\mathscr{B}_{m-1}$, as a single rank-one 
IC-POVM, $\{F(x)=\tau(x)\pi(x)\}_{x\in\mathscr{X}}$ say, by setting $\mathscr{X}=\cup_a\mathscr{B}_a$ and 
$\tau(x)=\sum_a v_a 1_{\mathscr{B}_a}(x)$ for some choice of weights $v_a$, then the error in the tomographic process is 
minimized if and only if the outcome set $\mathscr{X}$ forms a weighted $2$-design with weight $w(x)=\tau(x)/d$, and 
Eq.~(\ref{eq:reconstruct}) is used for state reconstruction. As we remarked in Sec.~\ref{subsec:qse}, one way of realizing 
the IC-POVM is to perform random orthogonal measurements as specified by the bases. In this case basis $\mathscr{B}_a$ is chosen 
with probability $v_a$. Smaller error rates are achieved, however, if we instead choose $\mathscr{B}_a$ exactly $v_a N$ times. 
The following subsection focuses on this latter scenario. 

\subsection{Optimal quantum state tomography by a series of orthogonal measurements}
\label{subsec:qst}

When quantum state tomography is achieved through a series of orthogonal measurements, each specified by a member of a complete set of MUBs, 
it is customary to cycle through the bases in turn rather than select them randomly with equal probability. We will now treat this 
important scenario for a weighted complex projective 2-design composed of $m$ orthonormal bases, 
$\mathscr{X}=\cup_a\mathscr{B}_a$. Although our results will apply in the general case, where there might be a multiplicity in elements 
across different bases, for the sake of notational simplicity, we will assume in this subsection that this is not the case. 
That is, we will assume that $|\mathscr{X}|=md$, so that $w(e^a_j)=w_a$ (or $\tau(e^a_j)=v_a$) under the association 
$\mathscr{B}_a=\{e^a_j\}_{j=0}^{d-1}$.

For complete state determination we require that the bases are in fact eigenbases, prescribed by members of an informationally complete set 
of quantum observables~\cite{Prugovecki77,Schroeck89,Busch89,Busch91}. Alternatively, the bases must together form a rank-one IC-POVM, 
$\{F(e^a_j)=v_a\pi(e^a_j)\}_{a,j}$, for some choice of the positive weights $v_a$. We prefer this latter setting since state 
reconstruction now follows from the results of Sec.~\ref{subsec:qstg}. Note that if the POVM is informationally complete for one 
choice of the weights, then it is informationally complete for all choices. Thus, with the exception that $v_a>0$ and $\sum_a v_a=1$, 
the weights can be chosen arbitrarily. There will be one particular choice, however, which will ease the analysis. With these considerations in mind, let us now begin.

Suppose that we make a series of orthogonal measurements as specified by a family of bases $\mathscr{B}_0,\dots,\mathscr{B}_{m-1}$, 
which together form a rank-one IC-POVM, $\{F(e^a_j)=v_a\pi(e^a_j)\}_{a,j}$, for some choice of the weights $v_a>0$. 
Let $n_a$ denote the number of measurements in basis $\mathscr{B}_a$, and let $N \coloneq \sum_a n_a$ be the total number of measurements.
Since we know that the IC-POVM in fact describes a series of orthogonal measurements, we will replace our previous estimate of the 
statistics [Eq.~(\ref{eq:pestimator})] with
\begin{equation}\label{eq:pestimator2}
\hat{p}(e^a_j)\;=\;\hat{p}(e^a_j;y_1,\dots,y_N) \;\coloneq\;\frac{v_a}{n_a}\sum_{k=1}^N\delta(e^a_j,y_k)\;,
\end{equation}
but retain Eq.~(\ref{eq:linearreconstructe}) for our estimate of $\rho$. 
Now following Ref.~\cite{Scott06}, we use this equation and Eq.~(\ref{eq:linearreconstruct}) together to rewrite the 
squared Hilbert-Schmidt distance as
\begin{align}
\|\rho-\hat{\rho}\|^2 & \;=\; \sum_{x,y\in\mathscr{X}}\big(p(x)-\hat{p}(x)\big)\big(p(y)-\hat{p}(y)\big)\BraKetb{Q(x)}{Q(y)} \\
& \;=\; \sum_{a,b,j,k} \big(p(e^a_j)-\hat{p}(e^a_j)\big)\big(p(e^b_k)-\hat{p}(e^b_k)\big)\BraKetb{Q(e^a_j)}{Q(e^b_k)}\;, \label{eq:notationalproblem}
\end{align}
giving an error in the form
\begin{align}
e^{(\{\mathscr{B}_a,v_a\},Q)}(\rho) & \;\coloneq\; \mathop{\textrm{\large E}}_{y_1,\dots,y_N}\Big[\,\|\rho-\hat{\rho}(y_1,\dots,y_N)\|^2\,\Big] \label{eq:errorforbases}\\
& \;=\; \sum_{a,j,k}\frac{1}{n_a}\big(v_a p(e^a_j)\delta_{jk}-p(e^a_j)p(e^a_k)\big)\BraKetb{Q(e^a_j)}{Q(e^a_k)}\;,\label{eq:errorforbases2}
\end{align}
since
\begin{align}
\mathop{\textrm{\large E}}_{y_1,\dots,y_N} \Big[\big(p(e^a_j)-\hat{p}(e^a_j)\big)\big(p(e^b_k)-\hat{p}(e^b_k)\big)\Big] 
&\;=\; v_a v_b \mathop{\textrm{\large E}}_{y_1,\dots,y_N} \Big[\big(q(e^a_j)-\hat{q}(e^a_j)\big)\big(q(e^b_k)-\hat{q}(e^b_k)\big)\Big] \\
&\;=\; \frac{\delta_{ab}v_a^2}{n_a}\Big(q(e^a_j)\delta_{jk}-q(e^a_j)q(e^a_k)\Big) \\
&\;=\; \frac{\delta_{ab}}{n_a}\Big(v_a p(e^a_j)\delta_{jk}-p(e^a_j)p(e^a_k)\Big)\;,
\end{align}
where $q(e^a_j)\coloneq\tr[\pi(e^a_j)\rho]=p(e^a_j)/v_a$ is the probability of result $e^a_j$ for the $a$-th orthogonal measurement, 
our estimate for this probability is $\hat{q}(e^a_j)\coloneq\hat{p}(e^a_j)/v_a$ [under Eq.~(\ref{eq:pestimator2})], and the 
expectation is an elementary calculation. 

Now suppose that the rank-one IC-POVM $\{F(e^a_j)=v_a\pi(e^a_j)\}_{a,j}$ 
is in fact a tight IC-POVM. Choosing $n_a=v_a N$ and $Q(e^a_j)=R(e^a_j)=(d+1)\pi(e^a_j)-I$ for the dual frame, 
we find that Eq.~(\ref{eq:errorforbases2}) simplifies to
\begin{equation}
e^{(\{\mathscr{B}_a,v_a\},R)}(\rho) \;=\; \frac{(d+1)^2}{N}\bigg(1-\sum_{a,j} \frac{p(e^a_j)^2}{v_a} \bigg)\;.
\end{equation}
But for a tight rank-one IC-POVM
\begin{align}
\sum_{a,j}\frac{p(e^a_j)^2}{v_a} &\;=\; \sum_{a,j} v_a \tr\big[\pi(e^a_j)\otimes\pi(e^a_j)\cdot\rho\otimes\rho\big] \\
&\;=\; \frac{2}{d+1}\tr\big[\Pisym^{(2)}\cdot\rho\otimes\rho\big] \\
&\;=\; \frac{1}{d+1}\Big(1+\tr(\rho^2)\Big) \;,
\end{align}
using Eq.~(\ref{eq:superisom1}) with $\tau(e^a_j)=v_a$, achieving the error rate
\begin{equation}\label{eq:errorfordesignbases}
e^{(\{\mathscr{B}_a,v_a\},R)}(\rho) \;=\; \frac{d+1}{N}\Big(d-\tr(\rho^2)\Big)  \;.
\end{equation}
Although the dominating contribution of $d^2/N$ remains the same, a comparison of the two error rates confirms a small improvement:
\begin{equation}
0\;\leq\; e^{(F,R)}(\rho)-e^{(\{\mathscr{B}_a,v_a\},R)}(\rho) \;=\; \frac{1}{N}\Big(d\tr(\rho^2)-1\Big)\;\leq\;\frac{d-1}{N} \;.
\end{equation}
The difference can be attributed to Eq.~(\ref{eq:errorforbases2}) 
(as compared to Eq.~(76) of Ref.~\cite{Scott06}), which takes into account that the bases are now being chosen nonrandomly.

We will now show that the improved error rate [Eq.~(\ref{eq:errorfordesignbases})] is in fact the minimum possible for any 
family of bases $\mathscr{B}_0,\dots,\mathscr{B}_{m-1}$, and for any choice of the reconstruction OVD $\{Q(e^a_j)\}_{a,j}$,
i.e., for any $Q$ satisfying the dual frame condition
\begin{equation}\label{eq:dualconstraint}
\sum_{a,j} \Ketb{Q(e^a_j)}\Brab{F(e^a_j)} \;=\; \sum_{a,j}v_a\Ketb{Q(e^a_j)}\Brab{P(e^a_j)} \;=\; \Is\;,
\end{equation}
where $P(e^a_j)=\pi(e^a_j)$. Let $\rho(\sigma,U) \coloneq U\sigma U^\dag$ for some fixed quantum state 
$\sigma\in\Qd$ and define the average error in the linear tomographic estimate of $\rho(\sigma,U)$ as
\begin{equation}
e_\mathrm{av}^{(\{\mathscr{B}_a,v_a\},Q)}(\sigma) \;\coloneq\; 
\int_{\mathrm{U}(d)}\d\muu(U)\: 
e^{(\{\mathscr{B}_a,v_a\},Q)}(\rho(\sigma,U))\;,
\end{equation}
using Eq.'s (\ref{eq:linearreconstruct}), (\ref{eq:linearreconstructe}), (\ref{eq:pestimator2}) and (\ref{eq:errorforbases}). 
The following is then the main result of this section.

\begin{thm}\label{thm:main}
Let $\mathscr{B}_0,\dots,\mathscr{B}_{m-1}\subset\C P^{d-1}$ be a family of orthonormal bases for $\C^d$ with the 
property that $\{F(e^a_j)=v_a\pi(e^a_j)\}_{a,j}$ is an IC-POVM, under the association $\mathscr{B}_a=\{e^a_j\}_{j=0}^{d-1}$, 
for some choice of the positive constants $v_0,\dots,v_{m-1}$. Then for any fixed quantum state $\sigma\in\Qd$,
the average error in the linear tomographic estimate of $\rho(\sigma,U)$ after $N=\sum_a n_a$ orthogonal measurements, 
with $n_a>0$ of those measurements in the basis $\mathscr{B}_a$, satisfies
\begin{equation}\label{eq:optimalstatetomographyviabases}
e_\mathrm{av}^{(\{\mathscr{B}_a,v_a\},Q)}(\sigma) \;\geq\; \frac{d+1}{N}\Big(d-\tr(\sigma^2)\Big) \;,
\end{equation}
for all reconstruction OVDs $\{Q(e^a_j)\}_{a,j}$. Furthermore, equality occurs if and only if $\{F(e^a_j)=v_a\pi(e^a_j)\}_{a,j}$ is a tight rank-one IC-POVM for the choice $v_a=n_a/N$, and, 
assuming this choice for state reconstruction, $Q(e^a_j)=(d+1)\pi(e^a_j)-I + D_a$, where $\{D_a\}_a$ is any set of operators satisfying $\sum_a v_a D_a = 0$.
\end{thm}

In particular, equality holds in Eq.~(\ref{eq:optimalstatetomographyviabases}) for a tight rank-one IC-POVM whenever $Q$ is the 
canonical dual frame, namely $Q(e^a_j)=R(e^a_j) = (d+1)\pi(e^a_j)-I$. 

This theorem is the analogue of Theorem~18 in Ref.~\cite{Scott06} and its proof will be similar. 
We first establish that the canonical dual frame with respect to the trace measure is optimal for state reconstruction. 
Unlike the scenario considered in the previous subsection, however, where the bases were sampled randomly, the canonical dual is 
no longer the unique optimum. These facts are a consequence of the following lemma (which is analogous to Lemma~16 of 
Ref.~\cite{Scott06}). Recall the general definition of the canonical dual frame $R$ [Eq.~(\ref{eq:canonicaldual})].

\begin{lem}\label{lem:optimaldualframe}
Let $\{F(e^a_j)=v_a\pi(e^a_j)\}_{a,j}$ be an IC-POVM with reconstruction OVD $\{Q(e^a_j)\}_{a,j}$, where each 
$\mathscr{B}_a=\{e^a_j\}_{j=0}^{d-1}$ is an orthonormal basis for $\C^d$. Then
\begin{equation}
\sum_{a,j,k} v_a \Big(\delta_{jk}-\frac{1}{d}\Big) \BraKetb{Q(e^a_j)}{Q(e^a_k)} \;\geq\;\sum_{a,j,k} v_a \Big(\delta_{jk}-\frac{1}{d}\Big) \BraKetb{R(e^a_j)}{R(e^a_k)} \;,
\end{equation}
with equality if and only if $Q(e_k^a) = R(e_k^a) + D_a$, where $\{D_a\}_a$ is any set of operators satisfying $\sum_a v_a D_a = 0$.
\end{lem}
\begin{proof}
By analogy with the dual frame condition [Eq.~(\ref{eq:dualconstraint})], first note that 
\begin{equation}
\sum_{a,j,k} v_a\Ketb{Q(e^a_j)}\Brab{P(e^a_k)} \;=\; \sum_{a,j}v_a\Ketb{Q(e^a_j)}\Brab{\,I\,} \;=\; \KetBra{I} \;,
\end{equation}
since $\sum_k P(e^a_k)=\sum_k\pi(e^a_k)=I$ and the left-right action of Eq.~(\ref{eq:dualconstraint}) on $\Ket{I}$ shows that
$\sum_{a,j} v_a Q(e^a_j)=I$. Now $Q$ and $R$ are both dual frames, so defining $D\coloneq Q-R$ we have
\begin{equation}
\sum_{a,j,k} v_a\Ketb{D(e^a_j)}\Brab{P(e^a_k)} \;=\; \KetBra{I}-\KetBra{I} \;=\; 0 \;,
\end{equation}
and since $\Ketb{R(e^a_k)}\coloneq\mathcal{F}^{-1}\Ketb{P(e^a_k)}$, 
\begin{equation}
\sum_{a,j,k} v_a\Ketb{D(e^a_j)}\Brab{R(e^a_k)} \;=\; 0 \;,
\end{equation}
which means
\begin{equation}
\sum_{a,j,k} v_a\BraKetb{D(e^a_j)}{R(e^a_k)} \;=\; 0 \;.
\end{equation}
This is the analogue of Eq.~(89) in the proof of Lemma 16, Ref.~\cite{Scott06}, which in the current scenario takes the form
\begin{equation}
\sum_{a,j} v_a\BraKetb{D(e^a_j)}{R(e^a_j)} \;=\; 0 \;.
\end{equation}
Combining, we obtain
\begin{equation}
\sum_{a,j,k} v_a \Big(\delta_{jk}-\frac{1}{d}\Big) \BraKetb{D(e^a_j)}{R(e^a_k)} \;=\; 0 \;.
\end{equation}
Using this relation and the inequality
\begin{equation}
\sum_{a,j,k} v_a \Big(\delta_{jk}-\frac{1}{d}\Big) \BraKetb{D(e^a_j)}{D(e^a_k)} \;=\;\sum_{a,j} v_a \BraKetb{C(e^a_j)}{C(e^a_j)} \;\geq\; 0 \;,
\end{equation}
where we have set $C(e^a_j)\coloneq D(e^a_j)-\frac{1}{d}\sum_k D(e^a_k)$, we obtain
\begin{align}
\sum_{a,j,k} v_a \Big(\delta_{jk}-\frac{1}{d}\Big) \BraKetb{Q(e^a_j)}{Q(e^a_k)} &\;=\; \sum_{a,j,k} v_a \Big(\delta_{jk}-\frac{1}{d}\Big) 
\Big[ \BraKetb{R(e^a_j)}{R(e^a_k)} +\BraKetb{R(e^a_j)}{D(e^a_k)} \nonumber\\ 
&\qquad\qquad\qquad\;\; +\BraKetb{D(e^a_j)}{R(e^a_k)}+\BraKetb{D(e^a_j)}{D(e^a_k)} \Big] \\
&\;=\; \sum_{a,j,k} v_a \Big(\delta_{jk}-\frac{1}{d}\Big) \Big[ \BraKetb{R(e^a_j)}{R(e^a_k)} +\BraKetb{D(e^a_j)}{D(e^a_k)} \Big] \qquad\;\\
& \;\geq\; \sum_{a,j,k} v_a \Big(\delta_{jk}-\frac{1}{d}\Big) \BraKetb{R(e^a_j)}{R(e^a_k)}  \;,
\end{align}
which is our desired result. Equality holds if and only if $C(e_j^a) = 0$ for all $j$ and $a$, or equivalently, 
$D(e_j^a) = D_a$, an operator which is independent of $j$. Both $R(e_k^a)$ and $Q(e_k^a) = R(e_k^a) + D_a$ must remain dual 
frames, however, so it is necessary that $\sum_{a,j}v_a \Ketb{P(e^a_j)}\Brab{D_a} = \sum_a v_a \Ketb{I}\Brab{D_a} = 0$, i.e. 
$\sum_a v_a D_a = 0$.
\end{proof}

The following technical result is also needed to prove Theorem~\ref{thm:main}.
Let $T\coloneq\sum_{j,k}\ket{e_j}\bra{e_k}\otimes\ket{e_k}\bra{e_j}$, which is called the ``swap'' 
since $T\ket{\psi}\otimes\ket{\phi}=\ket{\phi}\otimes\ket{\psi}$.

\begin{lem}\label{lem:integral}
Let $\mu$ be the Haar measure on $\mathrm{U}(d)$. Then
\begin{equation}\label{eq:integral}
\int_{\mathrm{U}(d)}\d\mu(U)\, \left[U\pi(e_j)U^\dag\right]\otimes\left[U\pi(e_k)U^\dag\right] \;=\; \frac{1-\delta_{jk}/d}{d^2-1}\,I\otimes I +\frac{\delta_{jk}-1/d}{d^2-1}\,T \;.
\end{equation}
\end{lem}
\begin{proof}
We have previously seen that [Eq.~(\ref{eq:symavg})]
\begin{equation}
\int_{\mathrm{U}(d)}\d\mu(U)\, \left[U\pi(e_j)U^\dag\right]^{\otimes 2} \;=\; \tbinom{d+1}{2}^{-1}\,\Pisym^{(2)} \;=\; \frac{I\otimes I + T}{d(d+1)} \;,
\end{equation}
which is Eq.~(\ref{eq:integral}) for $j=k$. Now write $I=\sum_j\pi(e_j)$ and expand
\begin{align}
I\otimes I &\;=\; \int_{\mathrm{U}(d)}\d\mu(U)\, \left[U I U^\dag\right]\otimes\left[U I U^\dag\right] \\
&\;=\; \int_{\mathrm{U}(d)}\d\mu(U)\,\sum_{j,k} \left[U\pi(e_j)U^\dag\right]\otimes\left[U\pi(e_k)U^\dag\right] \\
&\;=\; \int_{\mathrm{U}(d)}\d\mu(U)\; \Big(d\left[U\pi(e_j)U^\dag\right]^{\otimes 2} + d(d-1)\left[U\pi(e_j)U^\dag\right]\otimes\left[U\pi(e_k)U^\dag\right]\Big) \\
&\;=\; \frac{I\otimes I + T}{d+1} + d(d-1)\int_{\mathrm{U}(d)}\d\mu(U)\,\left[U\pi(e_j)U^\dag\right]\otimes\left[U\pi(e_k)U^\dag\right] \;,
\end{align}
assuming $j\neq k$ in the last two lines. Solving this equation gives Eq.~(\ref{eq:integral}) for $j\neq k$.
\end{proof}

\begin{proof}[Proof of Theorem~\ref{thm:main}]
First note that since the weights $v_a$ could always be absorbed into $Q$ in the RHS of 
Eq.~(\ref{eq:dualconstraint}), we are free to set $v_a=n_a/N$ without any loss of generality, and thus do so.
Now using Eq.~(\ref{eq:errorforbases2}) we have
\begin{align}
e_\mathrm{av}^{(\{\mathscr{B}_a,v_a\},Q)}(\sigma) & \;\coloneq\; \int_{\mathrm{U}(d)}\d\muu(U)\: e^{(\{\mathscr{B}_a,v_a\},Q)}(\rho(\sigma,U)) \\
& \;=\; \int_{\mathrm{U}(d)}\d\muu(U)\sum_{a,j,k}\frac{1}{n_a} \big(v_a p(e^a_j)\delta_{jk}-p(e^a_j)p(e^a_k)\big)\BraKetb{Q(e^a_j)}{Q(e^a_k)} \\
& \;=\; \sum_{a,j,k}\frac{v_a}{N}\int_{\mathrm{U}(d)}\d\muu(U)\: \Big(\tr[\pi(e^a_j)U\sigma U^\dag]\delta_{jk}-\tr[\pi(e^a_j)U\sigma U^\dag]\tr[\pi(e^a_k)U\sigma U^\dag]\Big) \nonumber\\
&\qquad\qquad\qquad\qquad\qquad\qquad\qquad\qquad\qquad \cdot\; \BraKetb{Q(e^a_j)}{Q(e^a_k)} \\
& \;=\; \sum_{a,j,k}\frac{v_a}{N} \bigg(\frac{\delta_{jk}}{d}-\int_{\mathrm{U}(d)}\d\muu(U)\:\tr\Big[[U^\dag\pi(e^a_j)U]\otimes[U^\dag\pi(e^a_k)U]\cdot\sigma\otimes\sigma\Big]\bigg) \nonumber\\
&\qquad\qquad\qquad\qquad\qquad\qquad\qquad\qquad\qquad \cdot\; \BraKetb{Q(e^a_j)}{Q(e^a_k)} \label{eq:mainprf1} 
\end{align}
since $p(e^a_j)\coloneq v_a\tr[\pi(e^a_j)\rho]= v_a\tr[\pi(e^a_j)U\sigma U^\dag]$ and 
$\int_{\mathrm{U}(d)}\d\muu(U) U\sigma U^\dag=I/d$. The remaining integral is the content of Lemma~\ref{lem:integral}, and since 
$\tr(T\cdot\sigma\otimes\sigma)=\tr(\sigma^2)$, Eq.~(\ref{eq:mainprf1}) reduces to
\begin{align}
e_\mathrm{av}^{(\{\mathscr{B}_a,v_a\},Q)}(\sigma) & \;=\; \frac{d-\tr(\sigma^2)}{(d^2-1)N} \sum_{a,j,k}v_a \Big(\delta_{jk}-\frac{1}{d}\Big)\BraKetb{Q(e^a_j)}{Q(e^a_k)} \\
& \;\geq\; \frac{d-\tr(\sigma^2)}{(d^2-1)N} \sum_{a,j,k}v_a \Big(\delta_{jk}-\frac{1}{d}\Big)\BraKetb{R(e^a_j)}{R(e^a_k)} \;,
\end{align}
applying Lemma~\ref{lem:optimaldualframe}. The sum can be simplified:
\begin{align}
\sum_{a,j,k}v_a \Big(\delta_{jk}-\frac{1}{d}\Big)\BraKetb{R(e^a_j)}{R(e^a_k)} & \;=\; \sum_a v_a \bigg(\sum_j\BraKetb{R(e^a_j)}{R(e^a_j)}-\frac{1}{d}\sum_{j,k}\BraKetb{R(e^a_j)}{R(e^a_k)}\bigg) \\ 
 & \;=\; \Big(\Tr(\mathcal{F}^{-1})-\frac{1}{d}\sum_{a,j,k} v_a \Brab{P(e^a_j)}\mathcal{F}^{-2}\Ketb{P(e^a_k)}\Big) \\ 
 & \;=\; \Big(\Tr(\mathcal{F}^{-1})-\frac{1}{d}\sum_a v_a \Brab{I}\mathcal{F}^{-2}\Ketb{I}\Big) \\ 
 & \;=\; \Big(\Tr(\mathcal{F}^{-1})-\frac{1}{d} \BraKetb{I}{I}\Big) \\ 
 & \;=\; \Big(\Tr(\mathcal{F}^{-1})-1\Big) \;,
\end{align}
using Eq.'s~(44) and (42) of Ref.~\cite{Scott06}, i.e. $\mathcal{F}^{-1}=\sum_{a,j}v_a\KetBrab{R(e^a_j)}$ and 
$\mathcal{F}\Ket{I}=\Ket{I}$, and also the fact that $\sum_j P(e^a_j)=\sum_j \pi(e^a_j)=I$. 

We have thus shown that
\begin{equation}
e_\mathrm{av}^{(\{\mathscr{B}_a,v_a\},Q)}(\sigma) \;\geq\; \frac{1}{(d^2-1)N} \Big(\Tr(\mathcal{F}^{-1})-1\Big)\Big(d-\tr(\sigma^2)\Big) \;,
\end{equation}
with equality if and only if $Q(e_k^a) = R(e_k^a) + D_a$, where $R$ is the canonical dual frame with respect to the trace measure $\tau(e^a_j)=v_a=n_a/N$, and $\sum_a v_a D_a = 0$. The remainder of the proof now follows from Lemma~17 of Ref.~\cite{Scott06}, which states that
\begin{equation}
\Tr\big(\mathcal{F}^{-1}\big) \;\geq\; d\big(d(d+1)-1\big) \;,
\end{equation}
with equality if and only if $\{F(e^a_j)=v_a\pi(e^a_j)\}_{a,j}$ is a tight rank-one IC-POVM.
\end{proof}

Finally, recalling that $\rho(\sigma,U) \coloneq  U\sigma U^\dag$ for some fixed $\sigma\in\Qd$, we define the 
worst-case error in the linear tomographic estimate of $\rho(\sigma,U)$ as  
\begin{equation}
e_\mathrm{wc}^{(\{\mathscr{B}_a,v_a\},Q)}(\sigma) \coloneq \sup_{U\in\mathrm{U}(d)}\: e^{(\{\mathscr{B}_a,v_a\},Q)}(\rho(\sigma,U)) \;,
\end{equation}
again using Eq.'s (\ref{eq:linearreconstruct}), (\ref{eq:linearreconstructe}), (\ref{eq:pestimator2}) and (\ref{eq:errorforbases}). 
The worst-case error is always bounded below by the average error, and when the bases form a tight rank-one IC-POVM, the error 
is in fact independent of $U$ [see Eq.~(\ref{eq:errorfordesignbases})]. We can thus immediately deduce the following corollary to 
Theorem~\ref{thm:main}.

\begin{cor}\label{cor:main}
Let $\mathscr{B}_0,\dots,\mathscr{B}_{m-1}\subset\C P^{d-1}$ be a family of orthonormal bases for $\C^d$ with the 
property that $\{F(e^a_j)=v_a\pi(e^a_j)\}_{a,j}$ is an IC-POVM, under the association $\mathscr{B}_a=\{e^a_j\}_{j=0}^{d-1}$, 
for some choice of the positive constants $v_0,\dots,v_{m-1}$. Then for any fixed quantum state $\sigma\in\Qd$,
the worst-case error in the linear tomographic estimate of $\rho(\sigma,U)$ after $N=\sum_a n_a$ orthogonal measurements, 
with $n_a>0$ of those measurements in the basis $\mathscr{B}_a$, satisfies
\begin{equation}
e_\mathrm{wc}^{(\{\mathscr{B}_a,v_a\},Q)}(\sigma) \;\geq\; \frac{d+1}{N}\Big(d-\tr(\sigma^2)\Big) \;,
\end{equation}
for all reconstruction OVDs $\{Q(e^a_j)\}_{a,j}$. Furthermore, equality occurs if and only if $\{F(e^a_j)=v_a\pi(e^a_j)\}_{a,j}$ is a tight rank-one IC-POVM for the choice $v_a=n_a/N$, and, 
assuming this choice for state reconstruction, $Q(e^a_j)=(d+1)\pi(e^a_j)-I + D_a$, where $\{D_a\}_a$ is any set of operators satisfying $\sum_a v_a D_a = 0$.
\end{cor}

In fact, weighted complex projective 2-designs which are composed of orthonormal bases specify the unique 
class IC-POVMs (describing a series of orthogonal measurements) that achieve
\begin{equation}
e_\mathrm{wc}^{(\{\mathscr{B}_a,v_a\},R)}(\sigma) \;=\; e_\mathrm{av}^{(\{\mathscr{B}_a,v_a\},R)}(\sigma) \;=\; e^{(\{\mathscr{B}_a,v_a\},R)}(\rho(\sigma,U))  \;=\; \frac{d+1}{N}\Big(d-\tr(\sigma^2)\Big) \;.
\end{equation} 

\section{Conclusion}
\label{sec:conclude}

In this article we have introduced the problem of constructing weighted 2-designs in $\C P^{d-1}$ from the union of a family of $m$ 
orthonormal bases for $\C^d$. If the weight remains constant across elements of the same basis, then such designs can be 
interpreted as generalizations of complete sets of MUBs, being equivalent whenever $m=d+1$ (Theorem~\ref{thm:MUBs}). Although 
weighted 2-designs can be constructed from orthonormal bases in all dimensions and for all sufficiently large $m\geq d+1$ 
(Corollary~\ref{cor:exist} and Theorem~\ref{thm:MUBs}), the task remains to find examples with $m$ as close as possible to the lower 
bound. To this end, we have presented explicit constructions of weighted 2-designs from $m=kd+2$ bases whenever $kd+1$ is a prime power, for any positive 
integer $k$ (Propositions~\ref{prp:npn} and \ref{prp:npngen} with Theorem~\ref{thm:npnbases}), and shown that $m=O(d^2)$ bases are 
always sufficient (Proposition~\ref{prp:quadfunc} with Theorem~\ref{thm:npnbases}). Furthermore, our approach, which is based on
highly nonlinear functions on abelian groups, sheds new light on the known constructions of complete sets of MUBs. Finally, we 
have shown that the entire class of weighted complex projective 2-designs which are composed of orthonormal 
bases specify the unique optimal choice of bases for quantum state tomography by orthogonal measurements (Theorem~\ref{thm:main} 
and Corollary~\ref{cor:main}).

Although this article was motivated from the practical standpoint of verifying quantum mechanical devices for information processing, 
quantum tomography provides one of the most powerful means to explore and test fundamental aspects of quantum theory. Indeed, 
quantum information processors rely so critically on the soundness of this theory that their very construction will 
provide new testament to its validity.

\begin{acknowledgments} 
The authors would like to thank Chris Godsil and Barry Sanders for their advice and input. AR is supported by NSERC and MITACS. AJS is supported by ARC and the State of Queensland.
\end{acknowledgments}

\end{document}